\documentclass[aps,prd,reprint,superscriptaddress,
  nofootinbib,floatfix,a4paper,amsmath,preprintnumbers]{revtex4-1}
\usepackage{graphicx} 
\usepackage{color}
\usepackage{mathptmx}
\usepackage{slashed}
\usepackage[sort&compress]{natbib}
\usepackage[colorlinks=true, linkcolor=blue, filecolor=blue,
  urlcolor=blue, citecolor=blue, plainpages=false]{hyperref}

\newcommand{\hmchipt}{HM$\chi$PT}
\def\mev{\,\mathrm{MeV}}
\def\gev{\,\mathrm{GeV}}
\def\fm{\,\mathrm{fm}}
\def\Tr{\,\mathrm{Tr}}

\newcommand\riken{RIKEN-BNL Research Center, Brookhaven National
  Laboratory, Upton, NY 11973, USA}
\newcommand\bnlaf{Physics Department, Brookhaven National Laboratory,
  Upton, NY 11973, USA}
\newcommand\Higgsaf{Higgs Centre for Theoretical Physics, School of
  Physics \& Astronomy, The University of Edinburgh, Edinburgh, EH9
  3FD, UK}
\newcommand\Juelichaf{Forschungszentrum J\"ulich, Institute for
  Advanced Simulation, J\"ulich Supercomputing Centre, 52425 J\"ulich,
  Germany}
\newcommand\sotonaf{School of Physics and Astronomy, University of
  Southampton, Southampton SO17 1BJ, UK}
\newcommand\uamaf{Instituto de F\'isica Te\'orica UAM/CSIC,
  Universidad Aut\'onoma de Madrid, E--28049 Madrid, Spain}

\begin{document}

\title{$B^*B\pi$ coupling using relativistic heavy quarks} 
\date{\today}
\author{J.M.~Flynn}
\affiliation\sotonaf
\author{P.~Fritzsch}\altaffiliation[Present address: ]{\uamaf}
\affiliation{Institut f\"ur Physik, Humboldt Universit\"at,
   12489 Berlin, Germany}
\author{T.~Kawanai}\altaffiliation[Present address: ]{\Juelichaf}
\affiliation\riken\affiliation\bnlaf
\author{C.~Lehner}
\affiliation{\bnlaf}
\author{C.T.~Sachrajda}
\author{B.~Samways} 
\affiliation\sotonaf
\author{R.S.~Van de Water}
\affiliation{Theoretical Physics Department, Fermi National
   Accelerator Laboratory, Batavia, IL 60510, USA}
\author{O.~Witzel}\altaffiliation[Present address: ]{\Higgsaf}
\affiliation{Center for Computational Science, Boston University,
   Boston, MA 02215, USA}

\collaboration{RBC and UKQCD Collaborations}

\preprint{FERMILAB-PUB-15-256-T}

\begin{abstract}
 We report on a calculation of the $B^*B\pi$ coupling in three-flavour
 lattice QCD. This coupling, defined from the strong-interaction
 matrix element $\langle B \pi | B^*\rangle$, is related to the
 leading order low-energy constant in heavy meson chiral perturbation
 theory (\hmchipt{}). We carry out our calculation directly at the
 $b$-quark mass using a non-perturbatively tuned clover action that
 controls discretization effects of order $|\vec{p}a|$ and $(ma)^n$
 for all $n$. Our analysis is performed on RBC/UKQCD gauge
 configurations using domain-wall fermions and the Iwasaki gauge
 action at two lattice spacings of $a^{-1}=1.729(25)\gev$,
 $a^{-1}=2.281(28)\gev$, and unitary pion masses down to $290\mev$. We
 achieve good statistical precision and control all systematic
 uncertainties, giving a final result for the coupling $g_b =
 0.56(3)_\mathrm{stat}(7)_\mathrm{sys}$ in the continuum and at the
 physical light-quark masses. This is the first calculation performed
 directly at the physical $b$-quark mass and lies in the region one
 would expect from carrying out an interpolation between previous
 results at the charm mass and at the static point.
\end{abstract}

\maketitle

\section{Introduction}
The power of lattice QCD in probing the Standard Model and uncovering
evidence for new physics lies predominantly in the flavour sector. To
constrain the Cabibbo--Kobayashi--Maskawa (CKM) unitarity
triangle~\cite{Charles2005, Bona2005, Laiho2010} requires many inputs
that must be evaluated non-perturbatively, particularly in the
$B$-meson sector. For instance, an important constraint on the apex of
the CKM unitarity triangle comes from neutral $B$-meson mixing, which
gives information on the ratio of CKM elements
$|V_{ts}|^2/|V_{td}|^2$. Accessing these CKM elements from the
experimental data requires knowledge of the $B$-meson decay constant
and bag parameter, or alternatively the SU(3) breaking ratio
\begin{equation}
   \label{eq:SU(3)breaking}
   \frac{f_{B_s}\sqrt{B_{B_s}}}{f_{B_d}\sqrt{B_{B_d}}}.
\end{equation}
Lattice calculations of the decay constants $f_{B_d}$ and $f_{B_s}$
are also necessary inputs for the Standard Model predictions of
BR$(B\to\tau\nu)$ and BR$(B_s\to\mu^+\mu^-)$ respectively, while
lattice calculations of the $B\to\pi l\nu$ form factor allow a
determination of the CKM matrix element $|V_{ub}|$. For both
semileptonic form factors and mixing matrix elements, the precision of
lattice calculations lags behind experiment. The experimental
measurements will continue to improve with the large data sets
available at Belle~II and from LHCb. Therefore it is essential to
reduce further the theoretical uncertainties in the non-perturbative
hadronic parameters in order to maximise the scientific impact of
current and future $B$-physics experiments.

A major source of uncertainty in all previous lattice calculations is
from practical difficulties simulating at physical light-quark masses.
Theoretical insight from \hmchipt{} can guide extrapolations down to
the physical point, but lack of knowledge of the low-energy constants
(LECs) of the theory introduces uncertainties. For example, at
next-to-leading order (NLO) in \hmchipt{} and lowest order in the
heavy-quark expansion the logarithmic dependence of $f_{B_d}$ and
$B_{B_d}$ on the light-quark (or equivalently, pion) mass is given
by~\cite{Goity:1992tp,Sharpe:1995qp}
\begin{align}
    f_{B_d} &= F\left(1-\frac34 \frac{1+3\hat g^2}{(4\pi f_\pi)^2}
    M_{\pi}^2\log(M_{\pi}^2/\mu^2)\right) + \cdots ,\label{eq:fB}
    \\ B_{B_d} &= B\left(1-\frac12 \frac{1-3\hat g^2}{(4\pi
      f_{\pi})^2} M_\pi^2\log(M_{\pi}^2/\mu^2)\right) + \cdots
    ,\label{eq:BB}
\end{align}
where $\hat g$ is the leading-order LEC. The strong-interaction matrix
element $\langle B\pi|B^*\rangle$ is used to determine a coupling
$g_b$, which would become $\hat g$ in the static limit of an
infinitely heavy $b$ quark. At the order used above in
Eqs.~\eqref{eq:fB} and~\eqref{eq:BB}, we are free to use $g_b$ in
place of $\hat g$; differences between the two are of order $1/m_b$.

In this paper we perform the first calculation of the coupling $g_b$
directly at the $b$-quark mass. Previous determinations of the
coupling have been hindered by the difficulties of simulating heavy
quarks on the lattice. Lattice calculations have been performed for
$g_c$, the analogous coupling for
$D$-mesons~\cite{Abada2002,Becirevic2004,Becirevic2012,Can2012}, and
for $\hat g$ itself~\cite{DeDivitiis1998, Becirevic2004,
  Ohki2008,Detmold2012,Bernardoni:2014kla}. Having a reliable
theoretical calculation of the coupling for the $B$ system is
important since this coupling cannot be accessed directly through
experiment. The strong coupling $g_{D^*D\pi}$ has been measured by the
CLEO collaboration~\cite{Ahmed:2001xc,Anastassov:2001cw} and more
recently by BaBar~\cite{Lees:2013zna,Lees:2013uxa}, but with
$B$-mesons there is not enough phase-space for the $B^*\rightarrow
B\pi$ decay to occur. Model estimates exist for $g_b$, including from
QCD sum
rules~\cite{Colangelo1994,Belyaev1995a,Dosch1996,Colangelo1998a} and
non-relativistic quark models~\cite{Yan1992a}.

The rest of this paper is organised as follows. In
section~\ref{heavymesontheory} we briefly review the framework of
HM$\chi$PT, show how $g_b$ enters and present the method for
extracting $g_b$ from lattice matrix-element calculations.
Section~\ref{computational_methods} describes the parameters of the
light-quark, gluon, and heavy-quark actions used in the numerical
calculation and presents the ratios of two- and three-point
correlators used to obtain $g_b$. In section~\ref{sec:analysis} we fit
the correlator ratios to extract $g_b$ and then extrapolate these
results to the continuum and physical quark masses using SU(2)
HM$\chi$PT. We estimate the systematic errors in $g_b$ in
section~\ref{sec:sys}, discussing each source of uncertainty in a
different subsection. We conclude in section~\ref{sec:conclusions} by
presenting our final results and error budget, and comparing our
result to other similar calculations.

\section{Heavy Meson Chiral Perturbation Theory}
\label{heavymesontheory}

In the infinite heavy-quark mass limit the properties of heavy-light
mesons become independent of the heavy quark's spin and flavour
quantum numbers. Combining this with the chiral symmetry present in
the massless light-quark ($m_q \rightarrow 0$) limit of QCD provides
the basis for heavy meson chiral perturbation theory. This effective
theory of QCD is a joint expansion in powers of the inverse
heavy-quark mass $1/m_Q$ and the light-quark-mass $m_q$.

In \hmchipt{} the heavy-light pseudoscalar and vector mesons, $P$ and
$P^*$, are combined in a covariant $4\times4$ matrix representation
\begin{equation}
  \label{eq:heavyfield}
  H = \frac{1+\slashed{v}}{2}
      \left(P^*_{\mu}\gamma^{\mu}-P\gamma_5\right).
\end{equation}
If one includes three light dynamical quark flavors ($u,d,s$), this
corresponds to SU(3) \hmchipt{} with the usual octet of
pseudo-Nambu-Goldstone bosons for the light pseudoscalars:
\begin{equation}
  \label{eq:pseudoscalaroctet}
  \mathcal{M} = \left(
    \begin{array}{ccc}
      \frac{1}{\sqrt{2}}\pi^0+\frac{1}{\sqrt{6}}\eta & \pi^+ & K^+ \\
      \pi^- & -\frac{1}{\sqrt{2}}\pi^0+\frac{1}{\sqrt{6}}\eta & K^0 \\
      K^- & \bar{K}^0 & -\sqrt{\frac{2}{3}}\eta
    \end{array}
    \right).
\end{equation}
Because the strange-quark mass is almost thirty times larger than the
average up-down quark mass, however, one can also treat the strange
quark as heavy and include only the up- and down-quark dynamical
degrees of freedom; this leads to SU(2) HM$\chi$PT (with the
corresponding modification of $\mathcal{M}$). At lowest order the
interactions between the heavy and light mesons are determined by a
Lagrangian with a single LEC~\cite{Wise1992,Casalbuoni1996}
\begin{equation}
  \label{eq:HMCHIPTLagrangian}
  \mathcal{L}_{\mathrm{HM}\chi\mathrm{PT}}^\mathrm{int} =
    \hat g \Tr (\bar{H}_aH_b\mathcal{A}_{ba}^\mu\gamma_\mu\gamma_5),
\end{equation}
where
\begin{equation}
  \mathcal{A}_{\mu} =
     \frac i2 \left(\xi^{\dagger}\partial_{\mu}\xi -
      \xi \partial_{\mu}\xi^{\dagger} \right)
\end{equation}
and $\xi = \exp(i\mathcal{M}/f_{\pi})$. The roman indices run over
light-quark flavour and the trace is over Dirac indices. We use a
convention where $f_\pi \approx 130\mev$.

The matrix element for the strong transition $B^* \rightarrow B\pi$ is
parametrised by $g_{B^*B\pi}$,
\begin{equation}
  \label{eq:gBstarBpi}
  \langle B(p') \pi(q)|B^*(p,\lambda)\rangle =
    g_{B^*B\pi}\; q\cdot\epsilon^{(\lambda)}(p),
\end{equation}
where $q=p-p'$ and $\epsilon^{(\lambda)}(p)$ is the polarization
vector for polarization state labelled by $\lambda$. Evaluating the
same matrix element at leading order in \hmchipt,
\begin{equation}
  \label{eq:gbstarbpihmchipt}
  \langle P(p') \pi(q)|P^*(p,\lambda)\rangle =
    \frac{2M_P}{f_\pi} \hat g \; q\cdot\epsilon^{(\lambda)}(p),
\end{equation}
enables the determination of $g_b$ from 
\begin{equation}
  \label{gbtogbstarbpi}
  g_{B^*B\pi} = \frac{2M_B}{f_{\pi}} g_b,
\end{equation}
with $g_b$ equal to $\hat g$ up to $1/m_b^n$ corrections.

Performing a Lehmann--Symanzik--Zimmermann reduction and using the
partially-conserved axial current relation for a soft pion,
Eq.~\eqref{eq:gBstarBpi} becomes
\begin{multline}\label{pionpole}
g_{B^*B \pi}\; q \cdot \epsilon^{(\lambda)}(p) =\\
    i q_{\mu} \frac{M_{\pi}^2-q^2}{f_{\pi}M_{\pi}^2}
     \int d^4x\; e^{iq\cdot x} 
    \langle B(p')|A^\mu(x)|B^*(p,\lambda)\rangle,
\end{multline}
where $A^\mu = \bar\psi_1 \gamma^\mu \gamma_5 \psi_2$ is the
light-quark axial-vector current. Using a form-factor decomposition of
the matrix element
\begin{multline}\label{formfactors} 
\langle B(p')| A^\mu | B^*(p,\lambda) \rangle =
  2M_{B^*} A_0(q^2) \,\frac{\epsilon^{(\lambda)}\cdot q}{q^2}q^\mu\\
  \begin{aligned}
  &+ (M_{B^*}+M_B)A_1(q^2)\left[\epsilon^{(\lambda)\mu} -
    \frac{\epsilon^{(\lambda)}\cdot q}{q^2}q^\mu\right] \\ 
  &+ A_2(q^2)\,\frac{\epsilon^{(\lambda)}\cdot q}{M_{B^*}{+}M_B}
    \left[p^\mu+p^{\prime\mu} - \frac{M_{B^*}^2{-}M_B^2}{q^2} q^\mu
    \right],
  \end{aligned}
\end{multline}
we see that at $q^2=0$
\begin{equation}
  g_{B^*B\pi}=\frac{2M_{B^*}A_0(0)}{f_{\pi}}.
\end{equation}
On the lattice, we cannot simulate exactly at $q^2=0$ without using
twisted boundary conditions. Furthermore and from
Eq.~\eqref{pionpole}, we see that the form factor $A_0$ contains a
pole at the pion mass, so it will be difficult to do a controlled
extrapolation to $q^2=0$. However, the decomposition in
Eq.~\eqref{formfactors} must be free of unphysical poles, which allows
us to obtain the relation
\begin{equation}
\label{gfromA1A2}
  g_{B^*B\pi} = \frac1{f_\pi}
     \left[ (M_{B^*}+M_B)A_1(0) + (M_{B^*}-M_B)A_2(0)\right].
\end{equation}
The $A_1$ term is expected to dominate because the relative
contribution of $A_2$ is suppressed by the ratio $(M_{B^*} -
M_B)/(M_{B^*} + M_B)$ whose value is $0.004$ for the physical $B$ and
$B^*$ masses. It is this relation that we use for our numerical
calculation.

\section{Calculational Strategy}
\label{computational_methods}

\subsection{Light quarks and gauge fields}

Our analysis is carried out using ensembles produced by the RBC and
UKQCD collaborations~\cite{Allton:2008pn, Aoki2010a} with the Iwasaki
gauge action~\cite{Iwasaki1983, Iwasaki1984} and $2{+}1$ flavour
dynamical domain-wall fermions~\cite{kaplan:1992bt,Shamir1993b}. The
configurations are at two lattice spacings, the finer $32^2$ ensembles
have an inverse lattice spacing of $a^{-1} = 2.281(28)\gev$ and the
coarser $24^3$ ensembles have $a^{-1} = 1.729(25)\gev$, corresponding
to approximately $0.086\fm$ and $0.11\fm$ respectively. All ensembles
have a spatial extent of $2.6\fm$. We simulate with unitary
light-quarks corresponding to pion masses down to $M_\pi=289\mev$. On
all ensembles the strange sea-quark mass is tuned to within 10\% of
its physical value. The fifth dimensional extent of both lattices is
$L_S = 16$, corresponding to a residual quark mass of
$m_\mathrm{res}a=0.003152$ on the $24^3$ ensembles and
$m_\mathrm{res}a=0.0006664$ on the $32^3$ ensembles. The lattice quark
masses corresponding to the physical $u/d$ and $s$ quarks are $\tilde
m_{u/d}a = 0.00136(4)$, $\tilde m_sa = 0.0379(11)$ on the $24^3$
ensembles and $\tilde m_{u/d}a = 0.00102(5)$, $\tilde m_sa =
0.0280(7)$ on the $32^3$ ensembles. The tildes indicate that these
values include the residual quark mass.

Our calculation makes use of unitary light-quark propagators with
point sources previously generated as part of the RBC/UKQCD
$B$-physics
program~\cite{Aoki:2012xaa,Witzel:2012pr,Christ:2014uea,Kawanai:2012id,Flynn:2015mha}.
Full details of the ensembles and propagators used are presented in
Table~\ref{fig:ensembles}. We perform a random translation on each
gauge field configuration to minimise the effects of autocorrelations
on our results, allowing us to use more closely spaced trajectories
and gain statistics. For each configuration in the $32^3$ ensembles we
use propagators computed at two time sources separated by half the
lattice temporal extent to compensate for the smaller ensemble sizes.
\begin{table}
  \caption{Lattice simulation parameters. All ensembles are generated
    using $2{+}1$ flavours of domain-wall fermions and the Iwasaki
    gauge action. All valence pion masses are equal to the sea-pion
    mass.}
  \begin{ruledtabular}
    \begin{tabular}[c]{ccccccc}
      $L/a$  & $a/\fm$ & $m_la$ & $m_sa$ & \#\ configs
         & \#\ sources & $M_{\pi}/\mev$\\
      \hline
      24 & 0.11  & 0.005 & 0.04  & 1636 & 1 & 329\\ 
      24 & 0.11  & 0.010 & 0.04  & 1419 & 1 & 419\\ 
      24 & 0.11  & 0.020 & 0.04  & 345 & 1 & 558\\ 
      32 & 0.08  & 0.004 & 0.03  & 628 & 2 & 289\\
      32 & 0.08  & 0.006 & 0.03  & 889 & 2 & 345\\
      32 & 0.08  & 0.008 & 0.03  & 544 & 2 & 394
    \end{tabular}
  \end{ruledtabular}
  \label{fig:ensembles}
\end{table}

\subsection{Bottom quarks}

Simulating heavy quarks on the lattice presents the problem of dealing
with $m_Qa\geq 1$. A number of approaches have been developed to
tackle this problem. In the limit of infinite mass the quarks become a
static source of colour charge and their lattice propagator reduces to
a trace of a product of temporal gauge links. This is the static
action of Eichten and Hill~\cite{Eichten1990} which has been used
extensively to calculate the coupling $g_\infty$, most recently
in~\cite{Detmold2012a,Bernardoni:2014kla}. Another approach is
nonrelativistic QCD~\cite{Lepage1992}, where the usual QCD Lagrangian
is expanded in powers of $v/c$.

Here we use the relativistic heavy-quark (RHQ)
action~\cite{El-Khadra1996a, Aoki2001, Christ2006} to simulate fully
relativistic bottom quarks while controlling discretization effects.
Although $m_Qa$ is large for the heavy quark in a heavy-light meson,
the spatial momentum $|\vec{p}a|$ is of $O(a \Lambda_\mathrm{QCD})$.
The RHQ action is an anisotropic Wilson action with a
Sheikholeslami-Wohlert term~\cite{Sheikholeslami1985}
\begin{widetext}
\begin{equation}
%\begin{multline}
\label{eq:RHQ}
S_\mathrm{RHQ} = a^4\sum_{x,y} \bar{\psi}(y)
   \bigg(m_0+\gamma_0D_0+\zeta\vec{\gamma}\cdot\vec{D}
   - \frac{a}{2}(D_0)^2-\frac{a}{2}\zeta(\vec{D})^2
   + \sum_{\mu \nu}\frac{ia}{4}c_p\sigma_{\mu\nu}F_{\mu\nu}
     \bigg)_{y,x}\psi(x).
%\end{multline}
\end{equation}
\end{widetext}
El-Khadra, Kronfeld and Mackenzie showed that, for correctly tuned
parameters, the anisotropic Clover action can be used to describe
heavy quarks with controlled cut-off effects to all orders in $m_Qa$
and of $O(|\vec{p}a|)$~\cite{El-Khadra1996a}. Christ, Li, and
Lin~\cite{Christ2006} later showed that only three independent
parameters need to be determined and, further, presented a method for
performing this parameter tuning non-perturbatively~\cite{Lin2006}.

This tuning has been completed for $b$ quarks~\cite{Aoki:2012xaa} on
the RBC/UKQCD configurations and those results
(Table~\ref{tab:RHQparams}) are exploited in this calculation. The
heavy-quark propagators and the correlation functions used in this
analysis are calculated using the \textsc{Chroma} software
library~\cite{Edwards:2004sx}. We apply Gaussian smearing to the
heavy-quark propagators using parameters tuned in~\cite{Aoki:2012xaa}.
Because the correlators become very small at large time separations
owing to the large masses in the exponentials, we run the inverter for
the heavy quark propagators to a very small relative residual
($10^{-45}$). We found that pursuing the conjugate gradient iteration
to this small residual is equivalent to demanding convergence
separately for the residual on each time slice.
\begin{table}
  \caption{Tuned RHQ parameters for $b$ quarks from
    Ref.~\cite{Aoki:2012xaa}. The uncertainties shown are statistical,
    heavy-quark discretization effects, lattice-scale uncertainty, and
    from experimental inputs (the spin-averaged $B_s$-meson mass and
    $B_s$ hyperfine splitting) respectively.}
  \begin{ruledtabular}
  \begin{tabular}[c]{cccc}
   $a/\fm$   & $m_0a$  & $c_p$  & $\zeta$  \\\hline
   0.11 & 8.45(6)(13)(50)(7) & 5.8(1)(4)(4)(2) & 3.10(7)(11)(9)(0) \\
   0.08& 3.99(3)(6)(18)(3)  & 3.57(7)(22)(19)(14)& 1.93(4)(7)(3)(0)
   \end{tabular}
  \end{ruledtabular}
   \label{tab:RHQparams}
\end{table}

\subsection{Three-point correlation functions}
\label{sec:correlation_functions}

\begin{figure}
\begin{center}
\includegraphics[width=\linewidth]{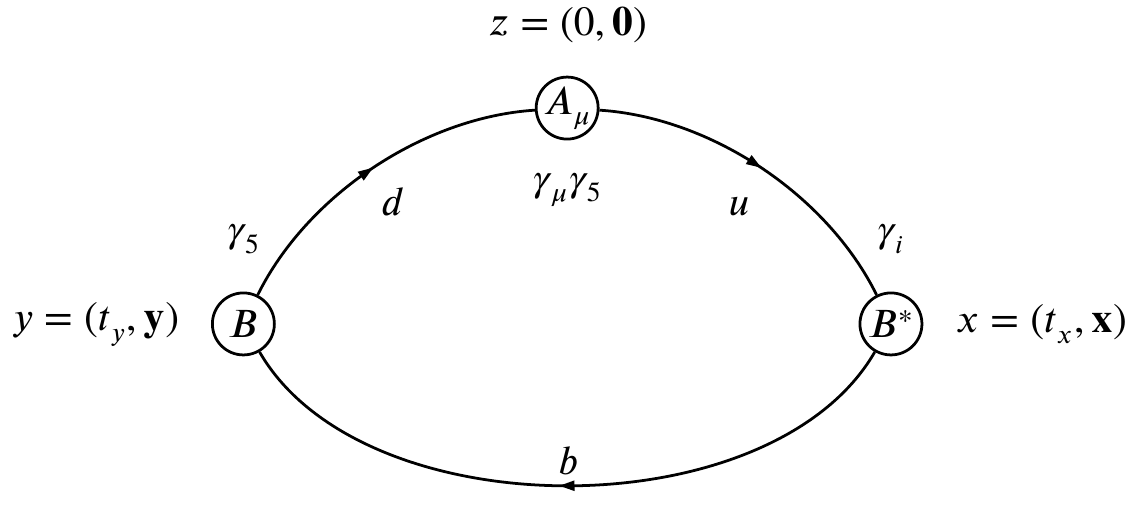}
\end{center}
  \caption{Quark flow diagram}
  \label{fig:quarkflow}
\end{figure}
\begin{figure*}
\begin{center}
\includegraphics[width=0.7\textwidth]{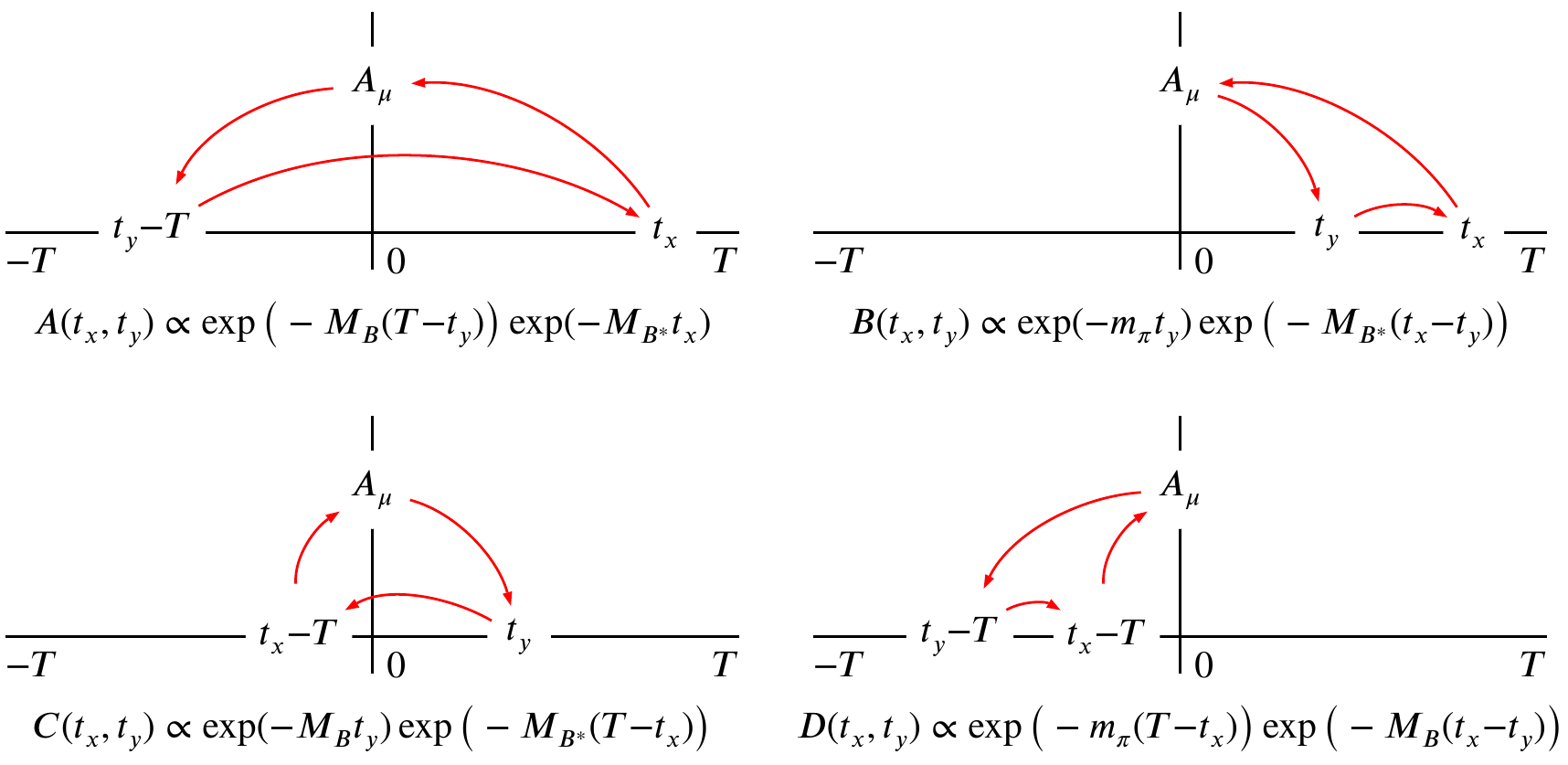}\\[3ex]
\includegraphics[width=0.7\textwidth]{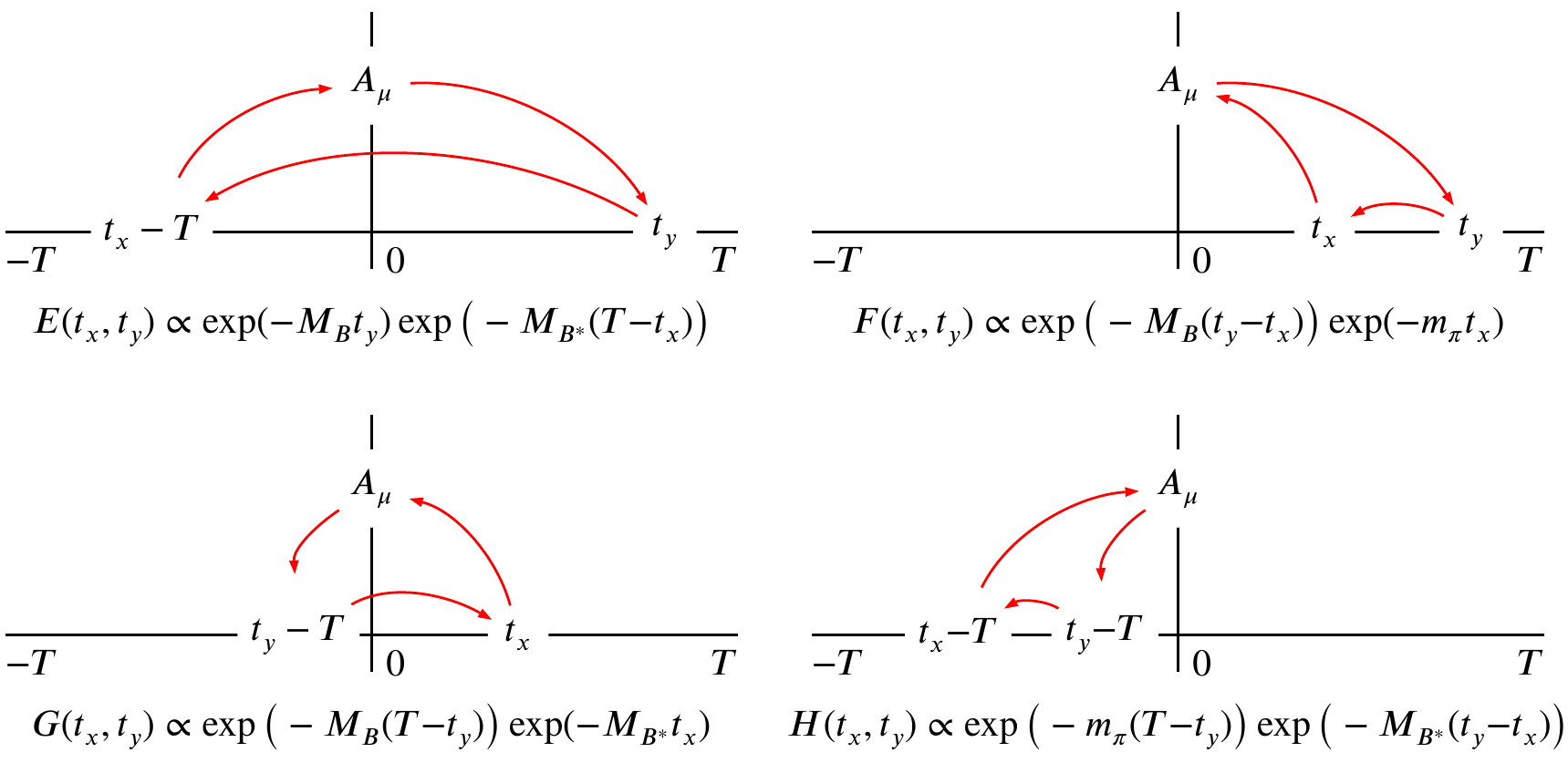}
\end{center}
  \caption{Leading contributions to the three-point correlator for
    $t_x > t_y$ ($A,B,C,D$) and $t_x\leq t_y$ ($E,F,G,H$).}
  \label{txty}
\end{figure*}
\begin{figure}
\begin{center}
    \includegraphics[width=0.95\linewidth]{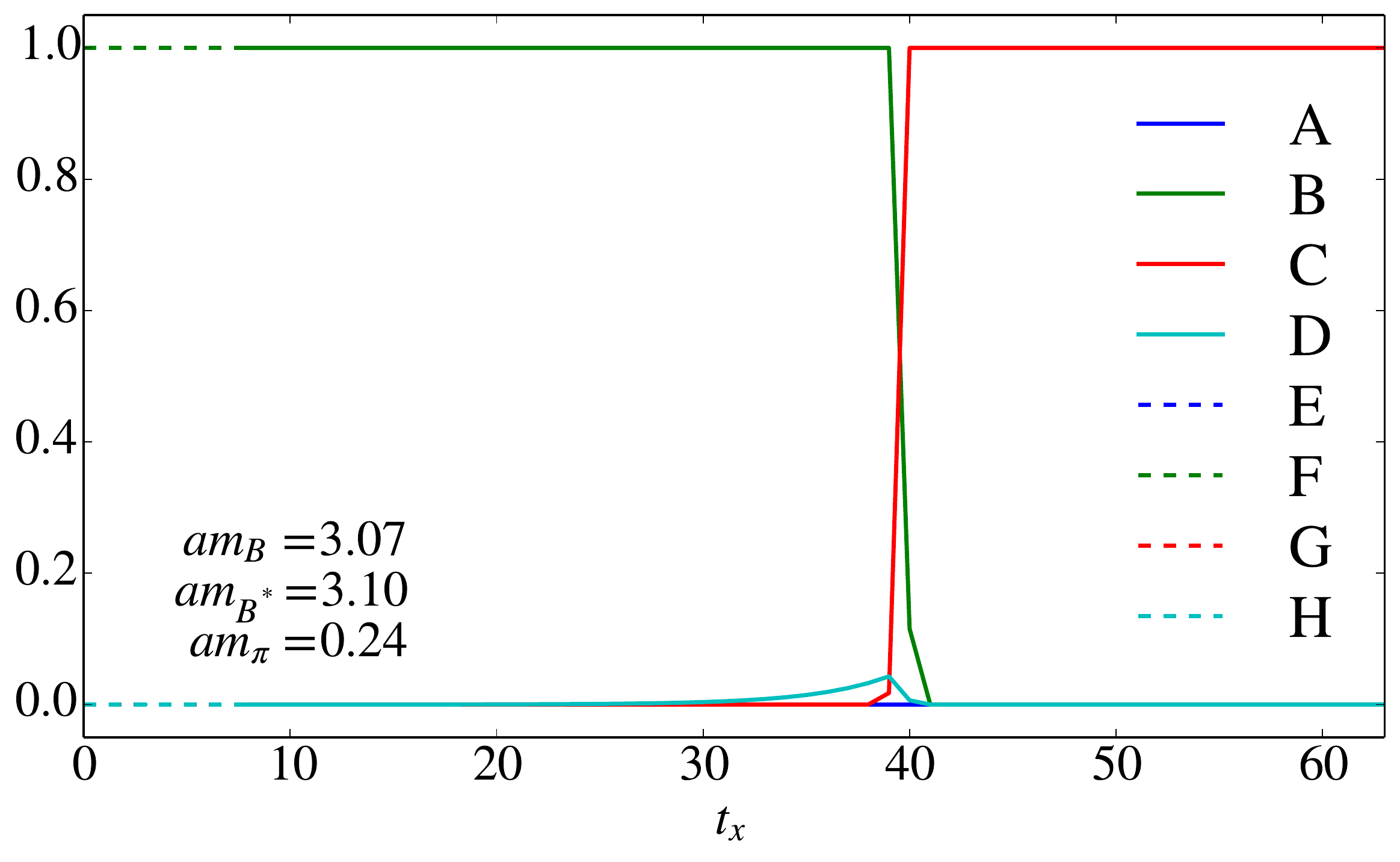}\\
    \includegraphics[width=0.95\linewidth]{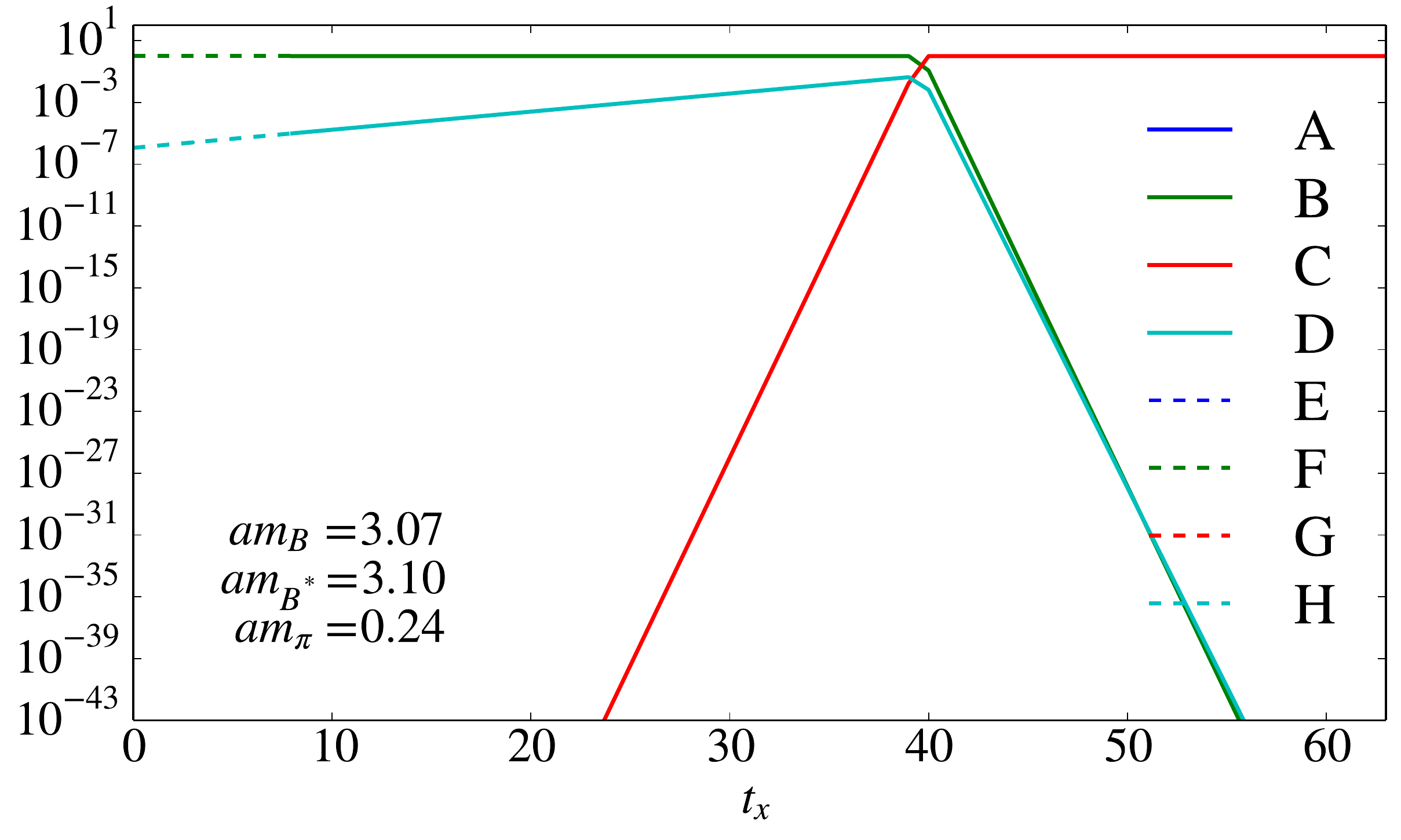}
\end{center}
\caption{The estimated relative sizes, with $t_y=8$, of the eight
  contributions to the $B^*B\pi$ three-point function arising from
  different Wick contractions, using a linear scale (top) and log
  scale (bottom). Each contribution is scaled by dividing by the
  maximum contribution at that time. The matrix element of interest
  comes from $C$ which is shown as a solid red line on both plots.
  This contribution dominates for large $t_x$.}
\label{fig:relative}
\end{figure}
\begin{figure*}
  \begin{center}
     \includegraphics[width=0.5\linewidth ]{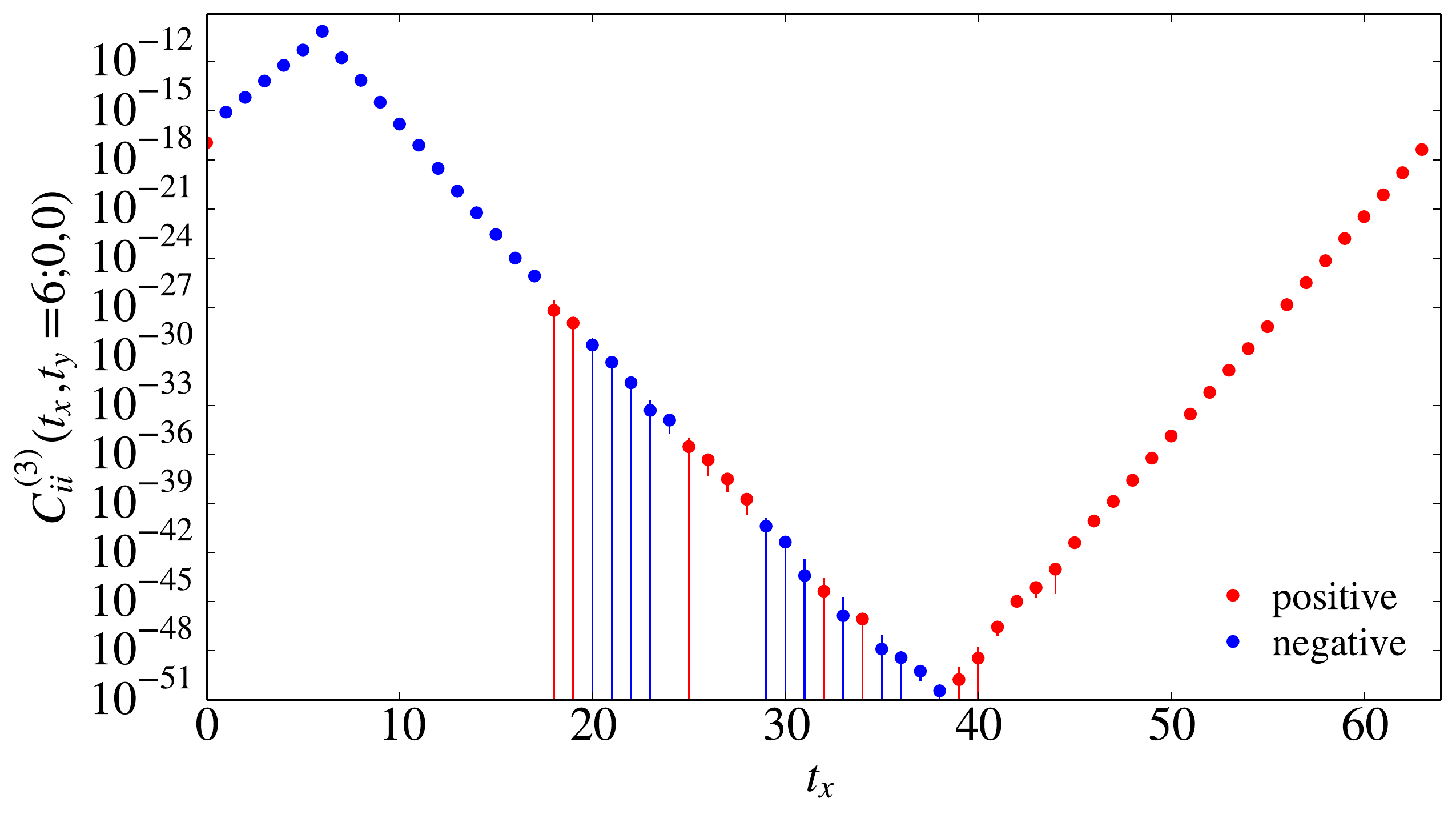}
     \includegraphics[width=0.491\linewidth ]{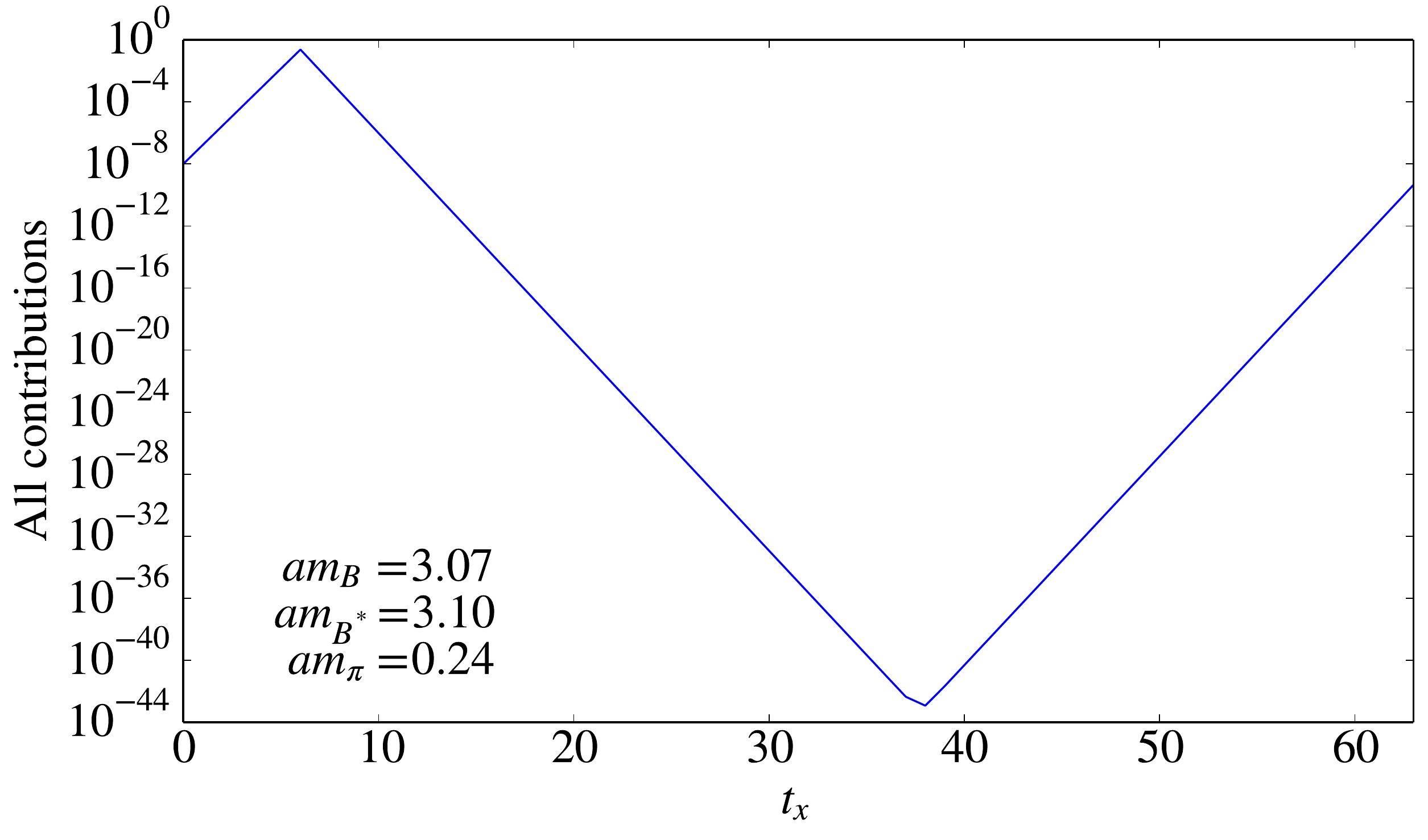}
  \end{center}
  \caption{The three-point correlator of Eq.~\eqref{threepoint}
    evaluated on the $24^3$, $m_la=0.005$ ensemble with $t_y=6$
    (left). The time dependence closely matches that predicted from
    the analysis in Sec.~\ref{sec:correlation_functions} (right).}
  \label{fig:three_point}
\end{figure*}

The matrix element that we wish to calculate in
Eq.~\eqref{formfactors} corresponds to the quark-flow diagram shown in
Fig.~\ref{fig:quarkflow}. To fully benefit from the available
pre-calculated light-quark propagators, we arrange the calculation so
that the axial-vector current is positioned at the light-quark
propagator's source. This means that we use the periodicity of the
lattice, creating the $B^*$ meson at large time $t_x$, propagating to
the origin at $t=0$, where the current acts and continuing to small
time $t_y$, where the $B$-meson is annihilated. With this setup, we
need only one inversion, a heavy-quark propagator calculated from
$t_y$, using a light-quark propagator for the sequential source. The
necessary trace is
\begin{multline}\label{trace}
  C_{\mu \nu}^{(3)}\left( t_x, t_y; \vec{p}, \vec{p}' \right) =\\
     \sum_{\vec{x},\vec{y}}
        e^{-i \vec{p}\cdot\vec{x}-i \vec{p}'\cdot\vec{y}}
     \,\mathrm{Tr}\left[
       S_l(0,y)\gamma_5S_h(y,x)\gamma_{\nu}
       S_l(x,0)\gamma_{\mu}\gamma_5
     \right].
\end{multline}
Because of the periodicity of the lattice there are further possible
contributions beyond the desired Wick contraction in
Fig.~\ref{fig:quarkflow}. An operator located at time $t$ can also be
considered at location $t-T$, where $T$ is the lattice's temporal
extent. If we take into account separately the cases where $t_x > t_y$
and $t_x \le t_y$, there are eight possible contributing arrangements
(considering $t+nT$ for integer $n$ gives further contributions,
but for increasing $|n|$ these are more and more suppressed). These
are shown in Fig.~\ref{txty}. Using the approximation that the ground
states immediately dominate the time dependence and that the matrix
elements are all unity, we can estimate the time dependence of the
three-point correlation function:
\begin{multline}
C^{(3)}(t_x, t_y) =\\ 
 \begin{cases}
  A(t_x, t_y)+B(t_x, t_y)+C(t_x, t_y)+D(t_x, t_y) & t_x > t_y,\\
  E(t_x, t_y)+F(t_x, t_y)+G(t_x, t_y)+H(t_x, t_y) & t_x \le t_y.
 \end{cases}
\label{eq:contributions}
\end{multline}
It is expected that the signal from which to extract the $B^*B\pi$
coupling will be seen at large $t_x$ (approaching $T$ from below),
coming from the contribution shown as $C(t_x,t_y)$ in Fig.~\ref{txty}.
Figure~\ref{fig:relative} shows the relative size of the different
contributions as a function of $t_x$ with all matrix elements set to
unity. As anticipated, $C$ is the dominant contribution in a region
$T/2+t_y < t_x < T$. This result appears steady for a range of the
masses $M_B, M_{B^*}, M_{\pi}$.
Plotting the sum of the contributions as a function of $t_x$ (setting
all matrix elements to unity) we see a peak at $t_x = t_y$ and an
overall cosh-like form shifted by $t_y$ as shown on the right in
Fig.~\ref{fig:three_point}. On the left of Fig.~\ref{fig:three_point}
we show good agreement with this form for our numerical data, with
$t_y$ the time-position of the source for the sequential inversion.
This gives us confidence that we can extract the desired matrix
element from the large $t_x$ region of the three-point correlator.


\subsection{Ratios}

To access the matrix element in Eq.~\eqref{formfactors} we calculate
the lattice three-point function
\begin{widetext}
\begin{equation}
    C_{\mu \nu}^{(3)}\left( t_x, t_y; \vec{p}, \vec{p}' \right) =
    \sum_{\vec{x},\vec{y}} e^{-i \vec{p}\cdot\vec{x}}
     e^{-i \vec{p}'\cdot\vec{y}} \langle B(y) A_\nu(0) B_\mu^*(x) \rangle
    \overset{\substack{T-t_x>0\\t_y>0}}\approx \sum_{\lambda}
      \frac{Z_B^{1/2}Z_{B^*}^{1/2}}{2E_B2E_{B^*}}
      \langle B(p')|A_{\nu}|B^*(p,\lambda)\rangle
      \epsilon^{(\lambda)}_{\mu} e^{-E_Bt_y-E_{B^*}(T-t_x)}
\label{threepoint}
\end{equation}
\end{widetext}
and the vector and pseudoscalar two-point functions
\begin{align}\label{twopoint}
C_{BB}^{(2)}\left(t;\vec{p}\right) &= \sum_{\vec{x}}
   e^{-i \vec{p} \cdot \vec{x} } \langle B(x) B(0)\rangle
   \approx Z_{B} \frac{e^{-E_{B}t}}{2E_{B}},\\
C_{B^*_{\mu}B^*_{\nu}}^{(2)}\left(t;\vec{p}\right) &= \sum_{\vec{x}}
  e^{-i \vec{p} \cdot \vec{x} } \langle B^*_{\nu}(x)
  B^*_{\mu}(0)\rangle\nonumber\\
   &\approx Z_{B^*}
  \frac{e^{-E_{B^*}t}}{2E_{B^*}}\left(\delta_{\mu \nu} -
  \frac{p_{\mu}p_{\nu}}{p^2}, \right).
\end{align}
If we set both the vector and pseudoscalar momenta to zero in
Eq.~\eqref{threepoint}, such that $\vec{q}=\vec{p}=\vec{p}'=0$ and
$q^2=q_0^2=(M_{B^*}-M_B)^2 \approx 0$, we can see from
Eq.~\eqref{formfactors} that the only form factor accessible is $A_1$.
Hence we form the ratio (not summed on $i$):
\begin{align}
  R_1(t_x,t_y) &= \frac{C_{ii}^{(3)}( t_x, t_y; \vec{p}=0,
    \vec{p}'=0 )\,Z_{B}^{1/2}Z_{B^*}^{1/2}}{C_{BB}^{(2)}
    (t_y;\vec{p}=0) \,C_{B^*_{i}B^*_{i}}^{(2)}
    (T-t_x;\vec{p}=0) }\nonumber\\
    & = (M_{B^*}+M_B)A_1(q_0^2),
    \label{R1def}
\end{align}
where we can average over the three spatial directions
($i=1,2,3$). To access the other form factors we need to inject a
unit of momentum, such that $\vec{q} = \vec{p} = (1,0,0)\times2\pi/L$
and $\vec{p}'=0$. Following~\cite{Abada2002}, we define the ratios:
\begin{align}
R_2(t_x,t_y) &= \frac{C_{10}^{(3)}( t_x, t_y;\vec{p}\neq0,
  \vec{p}'=0 )\,Z_{B}^{1/2}Z_{B^*}^{1/2}}{C_{BB}^{(2)}
  (t_y;\vec{p}'=0) \,C_{B^*_{2}B^*_{2}}^{(2)}
  (T-t_x;\vec{p}\neq0) }\nonumber\\
  &= \sum_{\lambda} \langle B(p') | A_0|B^*(p,\lambda)\rangle
           \epsilon_1^{(\lambda)},\\
R_3(t_x,t_y) &= \frac{C_{11}^{(3)}( t_x, t_y;\vec{p}\neq0,
   \vec{p}'=0 )\,Z_{B}^{1/2}Z_{B^*}^{1/2}}{C_{BB}^{(2)}
   (t_y;\vec{p}'=0) \,C_{B^*_{2}B^*_{2}}^{(2)}
   (T-t_x;\vec{p}\neq0) }\nonumber\\
   &= \sum_{\lambda} \langle B(p') | A_1|B^*(p,\lambda)\rangle
           \epsilon_1^{(\lambda)},\\
R_4(t_x,t_y) &= \frac{C_{22}^{(3)}( t_x, t_y;\vec{p}\neq0,
   \vec{p}'=0 )\,Z_{B}^{1/2}Z_{B^*}^{1/2}}{C_{BB}^{(2)}
   (t_y;\vec{p}'=0) \,C_{B^*_{2}B^*_{2}}^{(2)}
   (T-t_x;\vec{p}\neq0) }\nonumber\\
   &= \sum_{\lambda} \langle B(p') | A_2|B^*(p,\lambda)\rangle
            \epsilon_2^{(\lambda)}\nonumber\\
   &=(M_{B^*}+M_B)A_1(q^2).
\end{align}
These allow access to the form factor $A_2$ through
\begin{multline} \frac{A_2}{A_1} = \frac{(M_{B^*}+M_B)^2}{2M_B^2q_1^2}
   \Bigg[ -q^2_1+E_{B^*}(E_{B^*}-M_B)\\
          - \frac{M_{B^*}^2(E_{B^*}-M_B)}{E_{B^*}}\frac{R_3}{R_4}
          -i \frac{M_{B^*}^2q_1}{E_{B^*}}\frac{R_2}{R_4} \Bigg].
\label{A2/A1}
\end{multline}
The ratio in Eq.~\eqref{A2/A1} is obtained at non-zero values of $q^2$
and needs to be extrapolated to $q^2=0$. However, from
Eq.~\eqref{gfromA1A2} the contribution of $A_2(0)$ relative to
$A_1(0)$ is suppressed by the ratio $(M_{B^*} - M_B)/(M_{B^*} +
M_B)$. The form factor $A_1$ is obtained at $q_0^2 = (M_{B^*} -
M_B)^2$ from Eq.~\eqref{R1def}, but examination shows that the slight
extrapolation to $q^2=0$ is not necessary at the resolution possible
with the available statistics. If we define functions $G_1$ and $G_2$
\begin{equation}
  \label{eq:G1G2}
  \begin{split}
  G_1(q^2) &= (M_{B^*}+M_B)A_1(q^2),\\
  G_2(q^2) &= (M_{B^*}-M_B) A_2(q^2),
  \end{split}
\end{equation}
we can write the coupling as $G_1(0)$ plus a small correction from the
ratio $G_2/G_1$,
\begin{equation}
  \label{eq:gb}
  g_b = \frac{Z_A}{2M_B} G_1(0)\left( 1+\frac{G_2(0)}{G_1(0)} \right),
\end{equation}
where $Z_A$ is the light axial-vector current renormalization factor.
In our simulations $A_2$ is of comparable size to $A_1$. The mass
suppression in the ratio $G_2/G_1$ means that the correction term
in~\eqref{eq:gb} is at most $2\%$ on our ensembles and typically at
the sub percent level, with an error comparable to its size.

We take $Z_A$ from the RBC/UKQCD combined analysis of the light hadron
spectrum, pseudoscalar meson decay constants and quark masses on the
$24^3$ and $32^3$ ensembles~\cite{Aoki2010a}. The values are
calculated from the ratio of the conserved and local vector currents,
extrapolated to the chiral limit and are shown in
Table~\ref{table:renorm}.

\begin{table}
\caption{Axial current renormalization factors used in this work,
  calculated in~\cite{Aoki2010a}.}
\begin{ruledtabular}
\begin{tabular}[c]{ccc}
Ensemble & $a/\mathrm{fm}$ & $Z_A$\\
\hline
$24^3$ &  0.11  & 0.7019(26)\\
$32^3$ &  0.086 & 0.7396(17)\\
\end{tabular}
\end{ruledtabular}
\label{table:renorm}
\end{table}

\section{Analysis}
\label{sec:analysis}

\subsection{Correlator fits}

\begin{figure}
\begin{center}
\includegraphics[angle=0,width=\linewidth ]{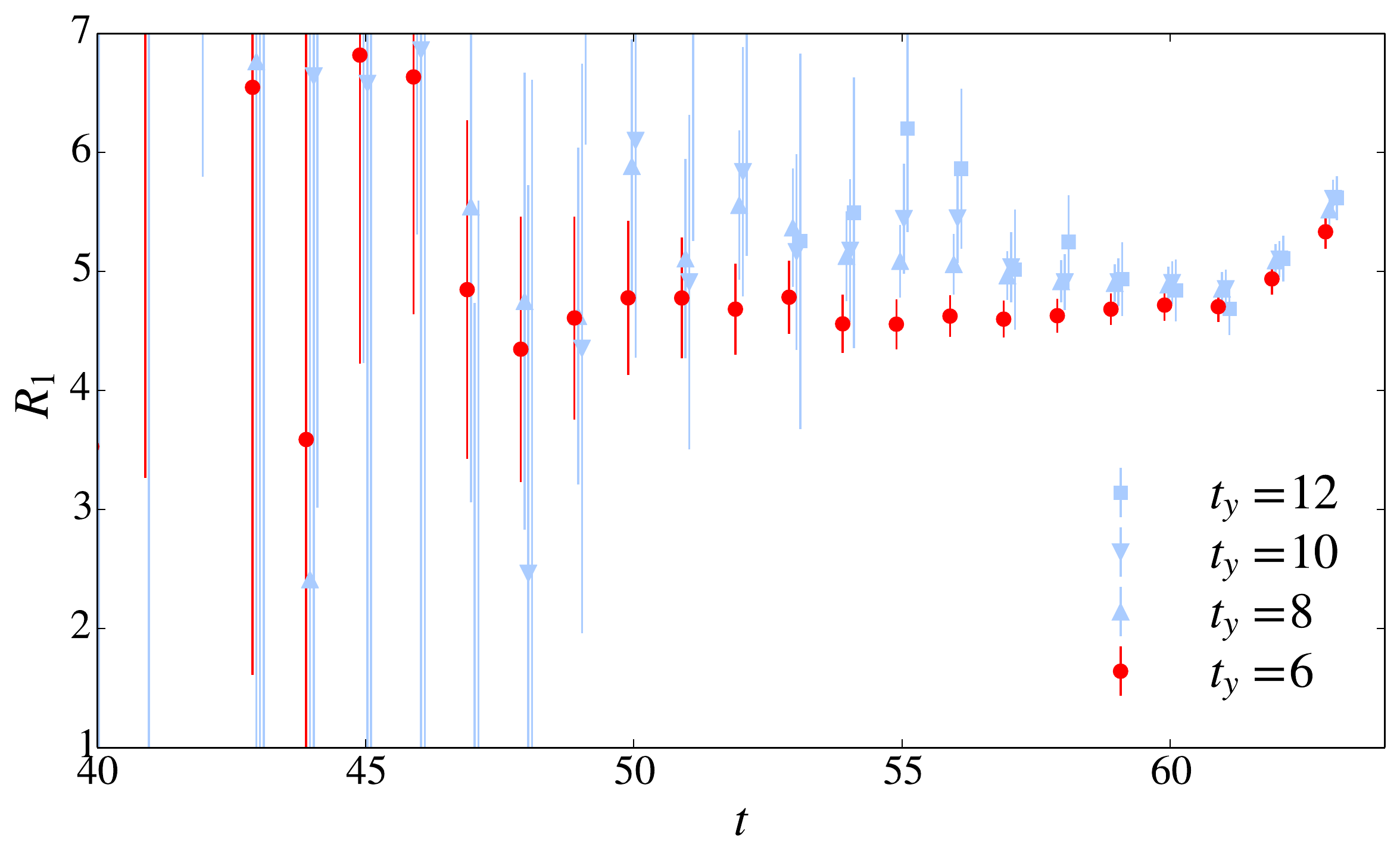}
\end{center}
\caption{The ratio $R_1(t,t_y)$ evaluated for different values of
  $t_y$ on the $24^3$, $m_la=0.005$ ensemble. $t_y=6$ gives the
  cleanest signal and longest plateau. The points for different $t_y$
  have small horizontal offsets to help distinguish them on the plot.}
  \label{fig:R1}
\end{figure}
We first calculate the three-point function on the $24^3$ ensemble
with $am_l = 0.005$ for values of $t_y$ ranging from $6$ to $18$.
Examining the data for $R_1$ (see Fig.~\ref{fig:R1}), it is clear that
the best signal is achieved with $t_y=6$. We therefore choose $t_y=6$
for our analysis on the $24^3$ ensembles and $t_y=8$ on the $32^3$
ensembles because it corresponds to the same physical distance.
\begin{figure*}
\begin{center}
\includegraphics[width=0.49\linewidth]{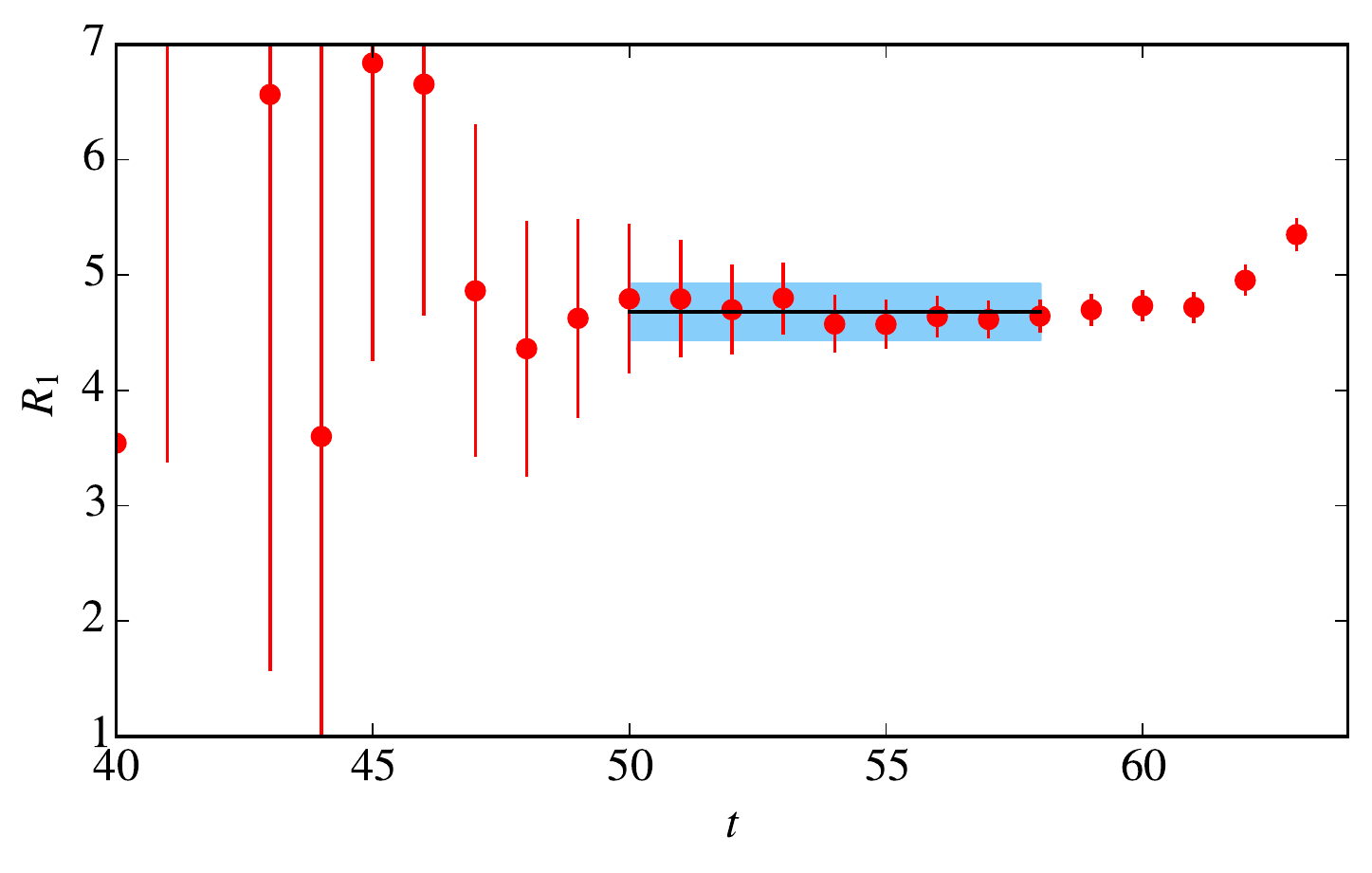}
\hspace*{\fill}
\includegraphics[width=0.49\linewidth]{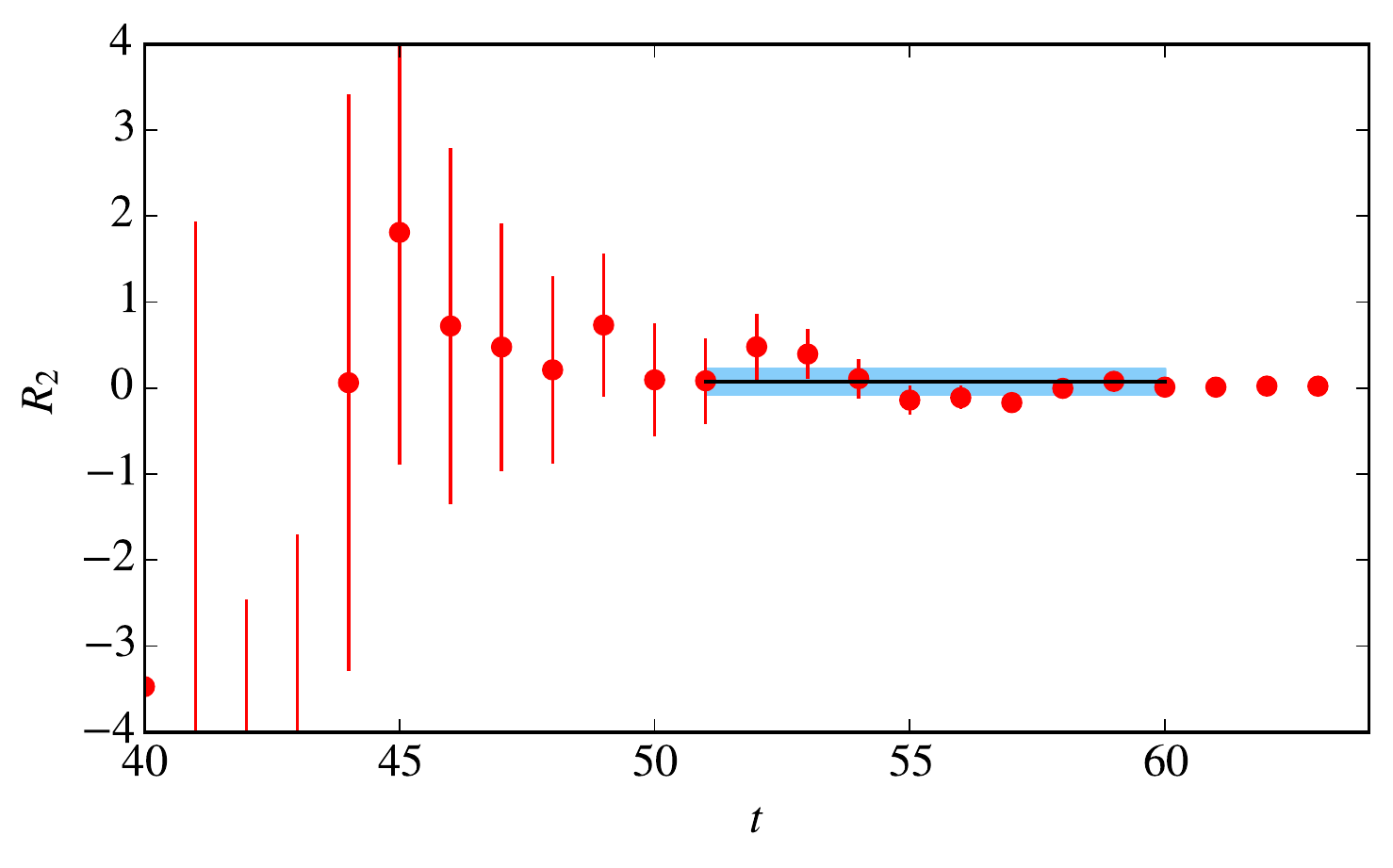}\\
\includegraphics[width=0.49\linewidth]{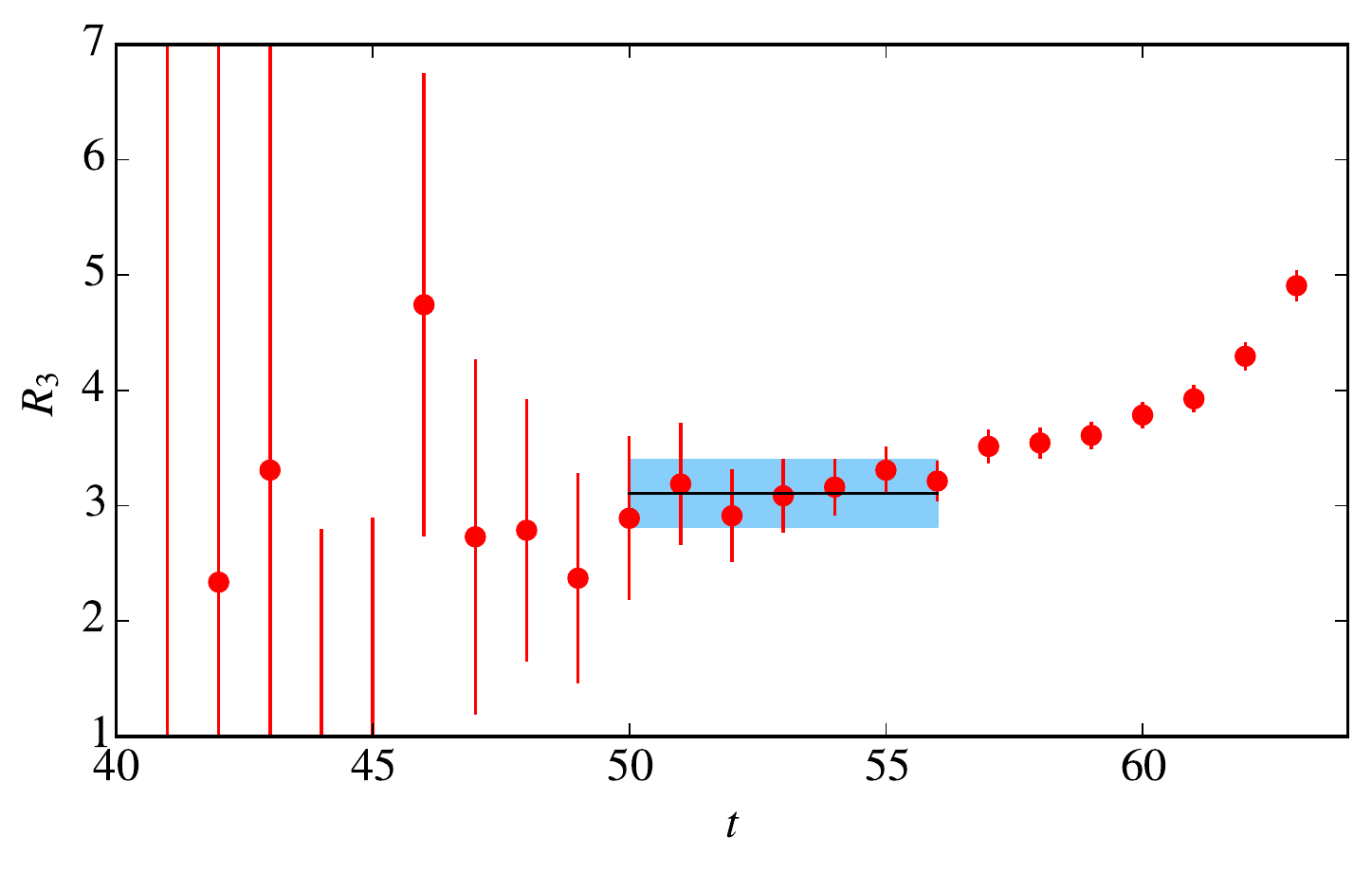}
\hspace*{\fill}
\includegraphics[width=0.49\linewidth]{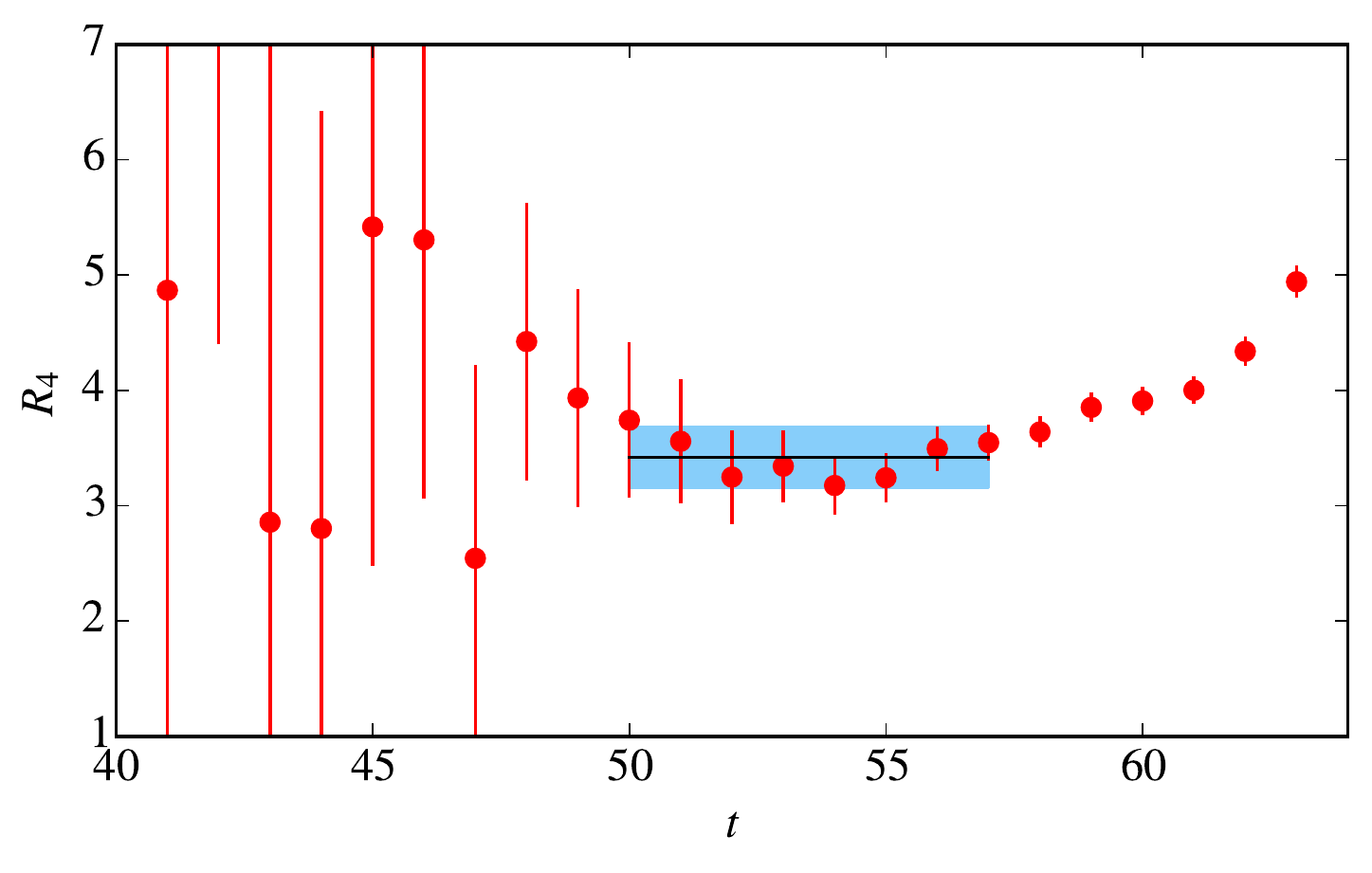}
\end{center}
  \caption{Ratios $R_1(t,t_y)$, $R_2(t,t_y)$, $R_3(t,t_y)$ and
    $R_4(t,t_y)$ for $t_y=6$ on the $24^3$, $m_la=0.005$ ensemble.
    $R_1$ is calculated with $\vec p = \vec p' = 0$, the other ratios
    with $\vec p=(1,0,0,0)2\pi/L$, $\vec p'=0$.}
  \label{fig:R1R2}
\end{figure*}

\begin{figure*}
\begin{center}
   \includegraphics[width=0.49\textwidth]{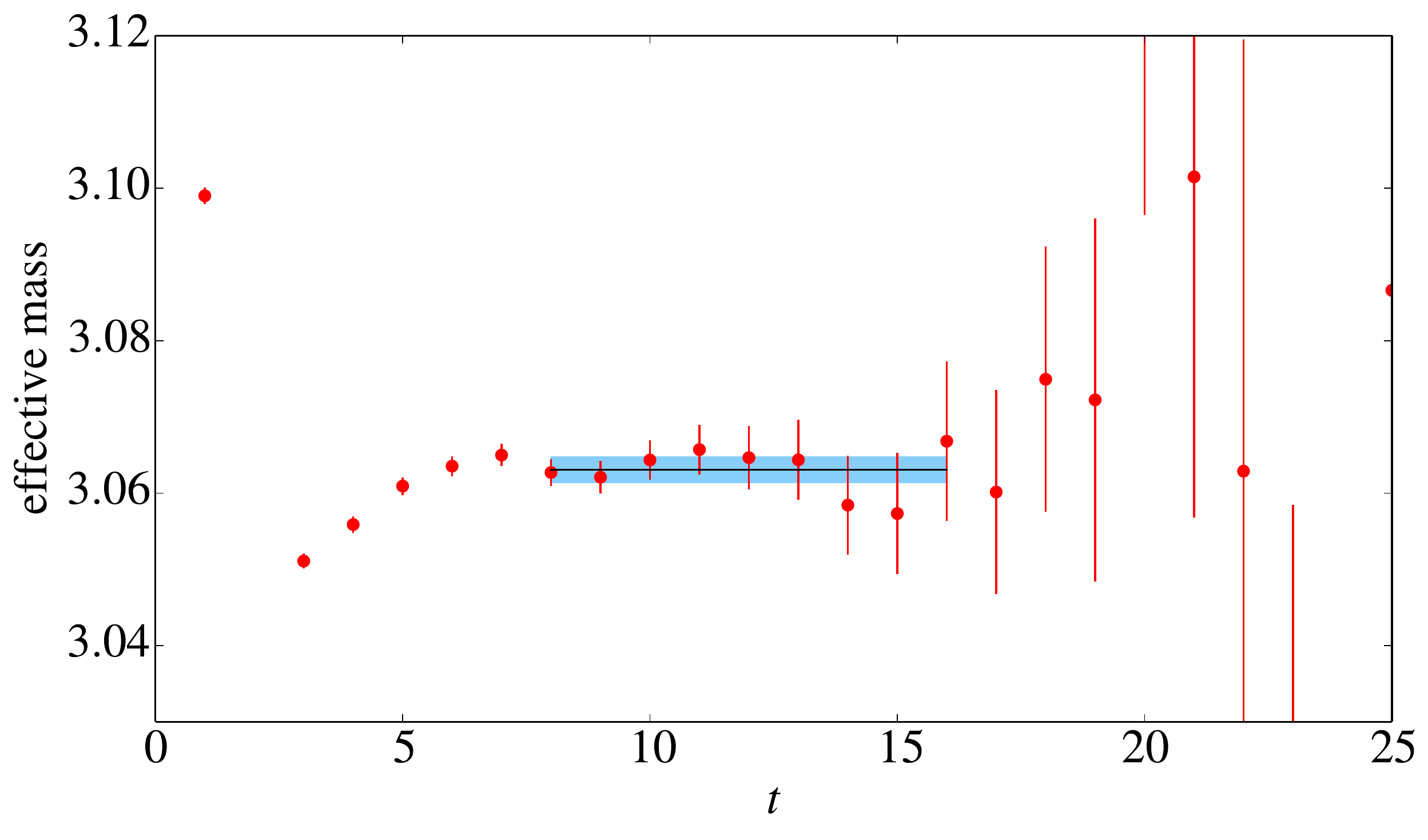}
   \hspace*{\fill}
   \includegraphics[width=0.49\textwidth]{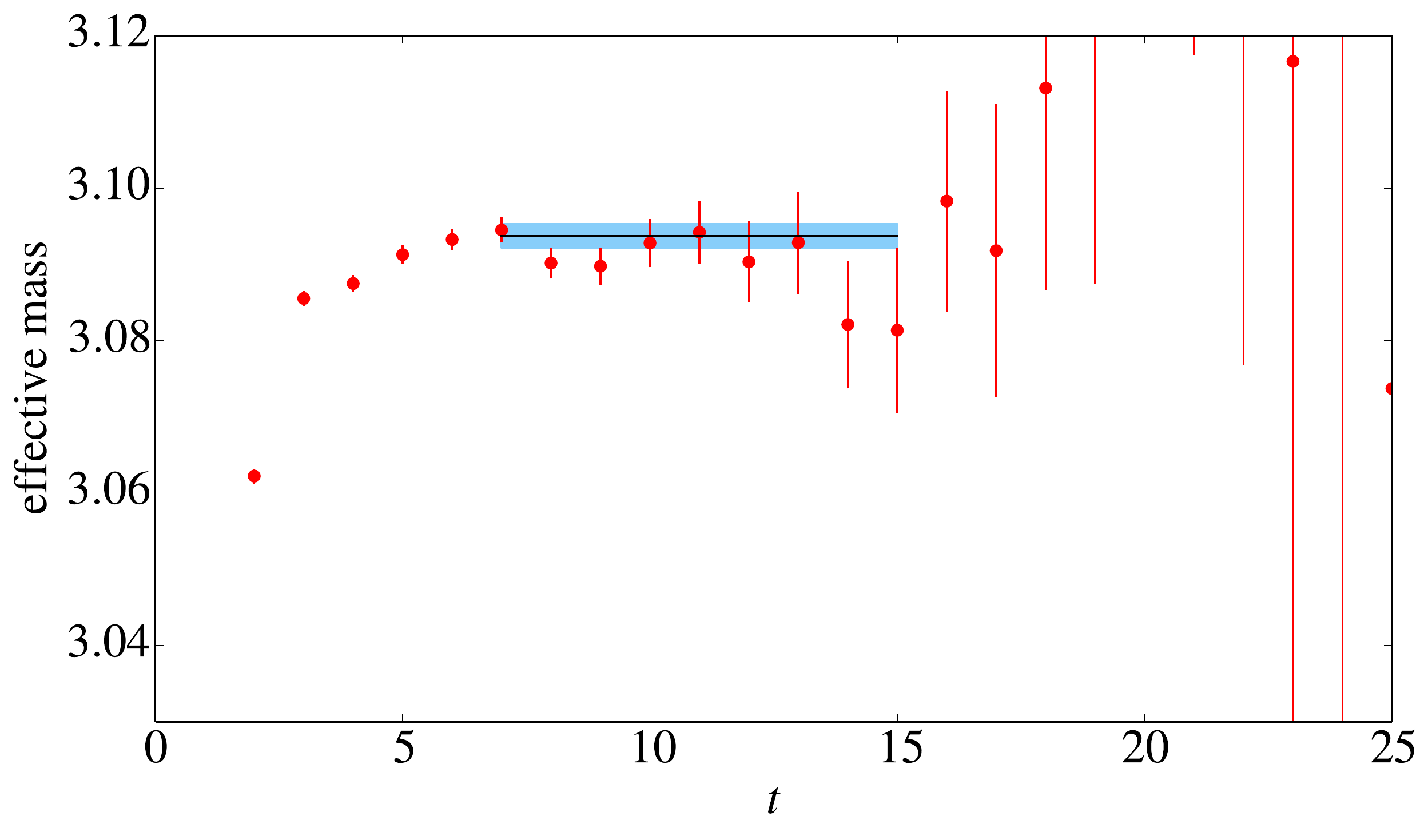}
\end{center}
\caption{$B$-meson (left) and $B^*$-meson (right) effective masses on
  the $24^3$, $m_la=0.005$ ensemble.}
\label{fig:PV}
\end{figure*}

Fig.~\ref{fig:R1R2} shows the ratios $R_1$, $R_2$, $R_3$ and $R_4$
calculated on the $24^3$, $m_la=0.005$ ensemble, and Fig.~\ref{fig:PV}
shows the vector and pseudoscalar effective-mass plots. In all cases,
we estimate the statistical error with a single-elimination jackknife.
The two-point functions are fitted to a single exponential to
extract $Z_{B}$ and $Z_{B^*}$. Using these values of $Z_{B^{(*)}}$, we
then fit the ratios to a constant in the regions given in
Table~\ref{fig:fit_ranges} where we expect excited-state contamination
to be small. We choose the fit ranges for each ratio such that we
obtain a good correlated $\chi^2/\mathrm{dof}$, and apply them to all
ensembles of the same lattice spacing consistently.

From the ratios we use the procedure described in the previous section
to extract $g_b$ on each ensemble, giving the values listed in
Table~\ref{tab:gbresults}.

\begin{table}
\caption{Fit ranges used for the two-point functions and the ratios.
  For non-zero momenta, equivalent combinations are averaged. Fit
  ranges are the same for different light-quark masses at the same
  lattice spacing, \emph{except} in the case of the $B^*$ two-point
  function with momentum $(1,1,0)$ on $24^3$, $m_la=0.020$ for which a
  range 7--15 is used in place of 9--15 listed in the table.}
\begin{ruledtabular}
\begin{tabular}{lcrr}
 & & \multicolumn{2}{c}{fit range $t_\mathrm{min}$--$t_\mathrm{max}$}\\
 \cline{3-4}
 & $\vec pa/(2\pi/L)$ & $24^3$ & $32^3$ \\
 \hline
$B$   & (0,0,0) & 8--16 & 8--17\\
$B^*$ & (0,0,0) & 7--15 & 8--16\\
$R_1$ & (0,0,0) & 50--58 & 50--58\\
$B^*$ & (1,0,0) & 7--15 & 8--16\\
$R_2$ & (1,0,0) & 51--60 & 47--55\\
$R_3$ & (1,0,0) & 50--56 & 46--55\\
$R_4$ & (1,0,0) & 50--57 & 47--56\\
$B^*$ & (1,1,0) & 9--15 & 10--16\\
$R_2$ & (1,1,0) & 51--60 & 47--55\\
$R_3$ & (1,1,0) & 49--55 & 46--55\\
$R_4$ & (1,1,0) & 51--57 & 46--54
\end{tabular}
\end{ruledtabular}
\label{fig:fit_ranges}
\end{table}

\begin{table}
\caption{Results for $g_b$ on the $24^3$ and $32^3$ ensembles. Errors
  for $g_b$ are statistical. Pion masses are
  from~\cite{Aoki2010a}.}
\begin{ruledtabular}
\begin{tabular}[c]{cccc}
$L/a$ & $m_la$ & $M_\pi^2/\gev^2$ & $g_b$ \\[0.5ex]\hline
$24$ & $0.005$ & $0.108$ & $0.533\pm0.027$ \\
$24$ & $0.010$ & $0.175$ & $0.568\pm0.023$ \\
$24$ & $0.020$ & $0.311$ & $0.580\pm0.035$ \\
$32$ & $0.004$ & $0.084$ & $0.548\pm0.020$ \\
$32$ & $0.006$ & $0.119$ & $0.603\pm0.016$ \\
$32$ & $0.008$ & $0.155$ & $0.596\pm0.018$
\end{tabular}
\end{ruledtabular}
\label{tab:gbresults}
\end{table}

\subsection{Chiral and Continuum Extrapolations}

We perform a chiral extrapolation using the SU(2) \hmchipt{} formula
for the axial coupling matrix element derived in~\cite{Detmold2011}:
\begin{equation}
  \label{eq:BecirevicChiralII}
  g_b = g_0\left(1 - \frac{ 2(1+2g_0^2) }{ (4\pi f_{\pi})^2 }
        M_{\pi}^2 \log \frac{ M_{\pi}^2 }{ \mu^2 } +
        \alpha M_{\pi}^2 + \beta a^2\right),
\end{equation} 
which is next-to-leading order in the chiral expansion, but only
leading order in the heavy-quark expansion. We parameterize the
light-quark and gluon discretization effects with an $a^2$ term, as
expected for the domain-wall light-quark and Iwasaki gauge actions.
The lattice-spacing dependence from the RHQ action is more
complicated. However, we argue in the next section that heavy-quark
discretization effects are negligible and that an extrapolation in
$a^2$ captures the leading scaling behaviour. We use the PDG
value~\cite{Beringer2012} of the pion decay constant,
$f_\pi=130.4\mev$.

\begin{figure}
  \centering
  \includegraphics[width=\linewidth ]{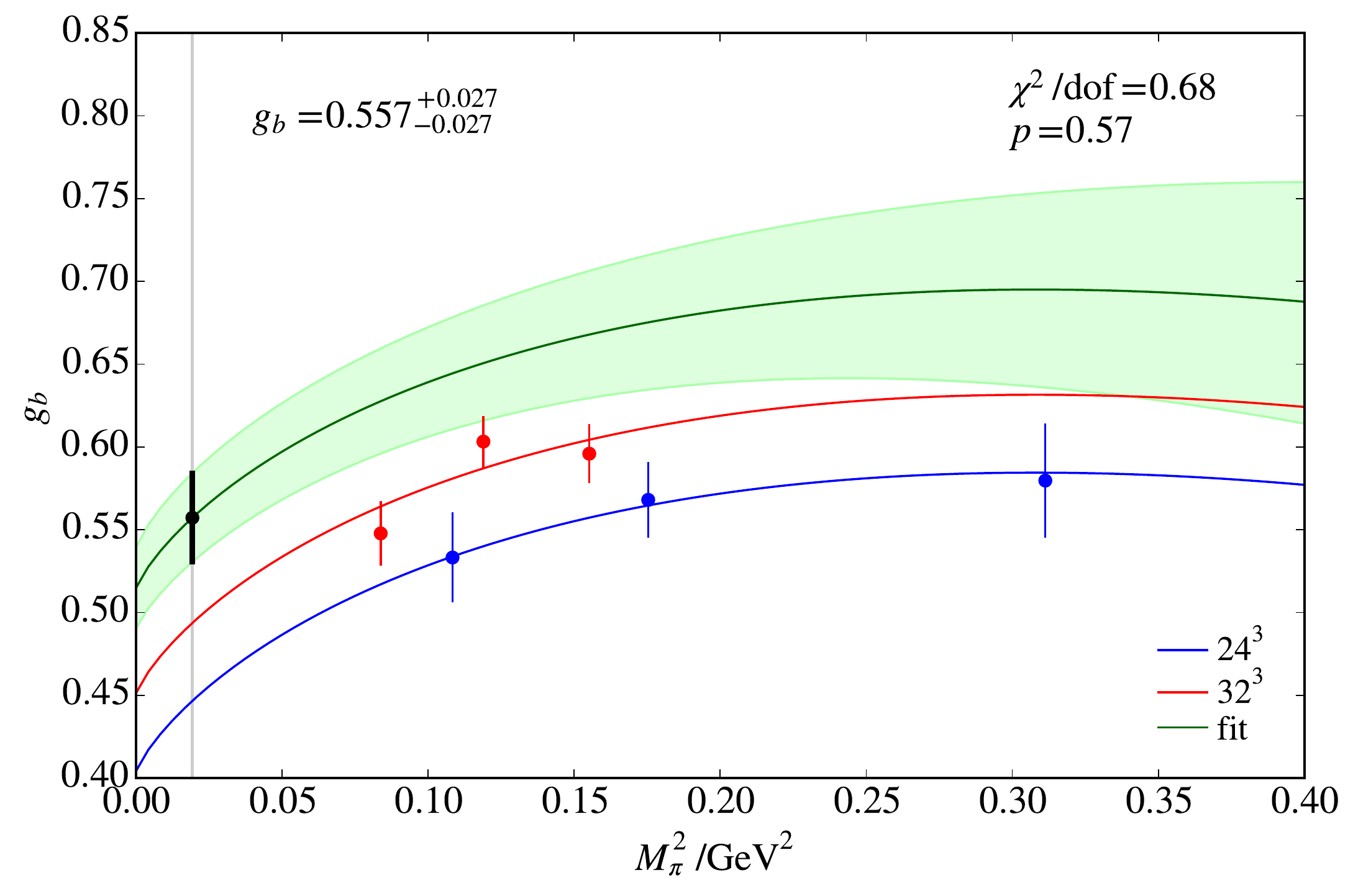}
  \caption{Chiral and continuum extrapolation. The bottom (blue) curve
    is the fit through the $24^3$ ensemble points. The (red) curve
    above is the fit through the $32^3$ ensemble points and the
    (green) error band and curve show the continuum extrapolation. The
    intersection with the vertical grey line corresponds to the
    physical pion mass. The result at the physical mass is shown by
    the (black) point with error.}\label{chiral_extrap}
\end{figure}

Fig.~\ref{chiral_extrap} shows the chiral-continuum extrapolation of
the numerical simulation data to the physical light-quark mass and
continuum using Eq.~\eqref{eq:BecirevicChiralII}. The values of $Z_A$,
as calculated in~\cite{Allton:2008pn} and~\cite{Aoki2010a}, are
included for each ensemble. The statistical errors in $Z_A$ are added
in quadrature to the Monte-Carlo statistical errors in the lattice
data for $g_b$ before performing the chiral fit. The six ensembles are
statistically independent, hence the fit is calculated by minimizing
an uncorrelated $\chi^2$ function. The parameters that minimise the
$\chi^2$ are $g_0=0.515$, $\alpha=-1.324\gev^{-2}$,
$\beta=-0.648\gev^2$. We use these fitted parameters and their
covariance matrix to estimate the error bands in our plots and to give
the errors in $g_b$. Our fitted result is $g_b = 0.557\pm0.027$.

\section{Systematic Errors}\label{sec:sys}

\subsection{Chiral Extrapolation}

\begin{figure*}
\begin{center}
% 1st row
\includegraphics[width=0.496\textwidth ]{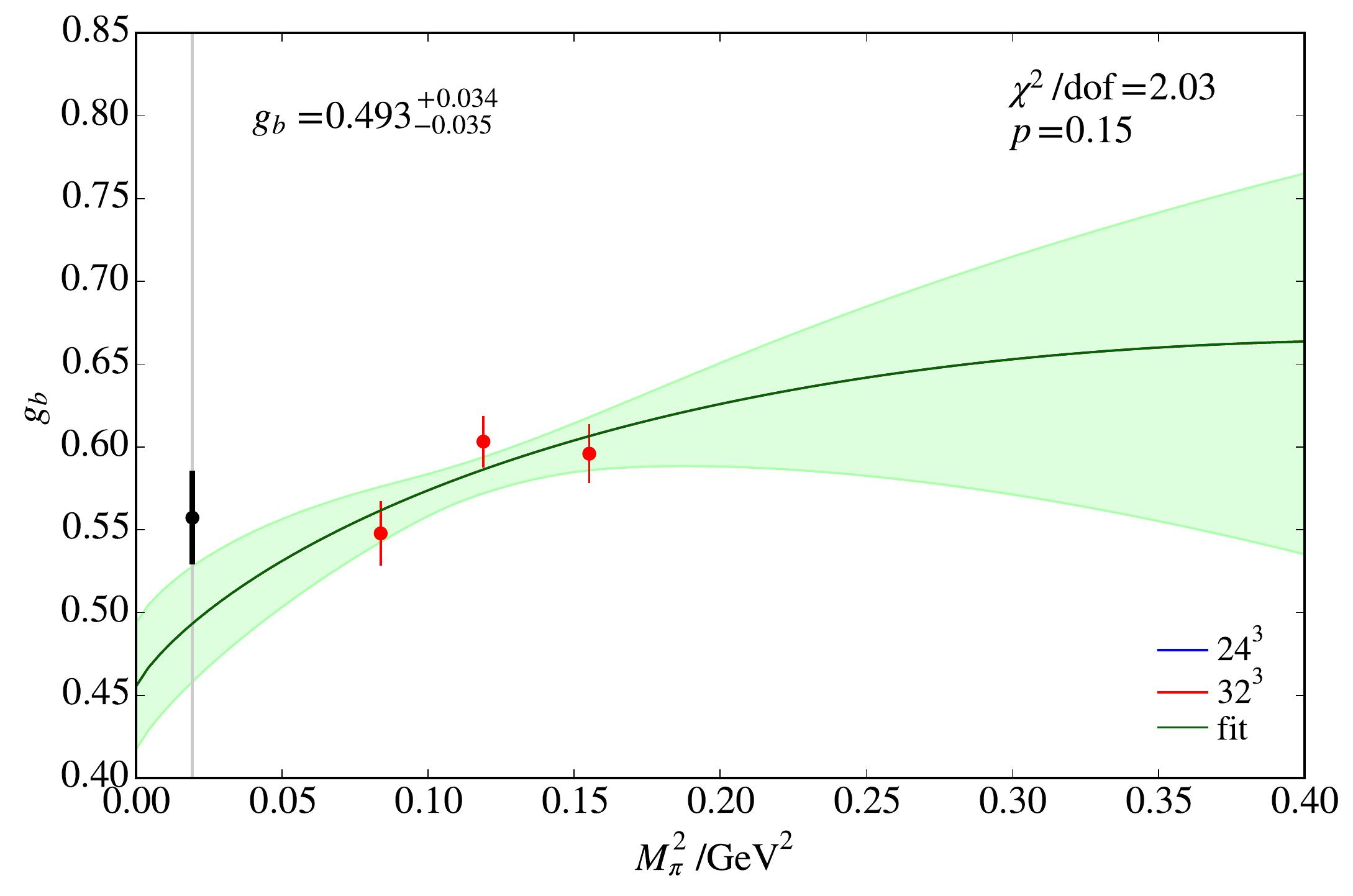}%
\hspace*{\fill}
\includegraphics[width=0.496\textwidth ]{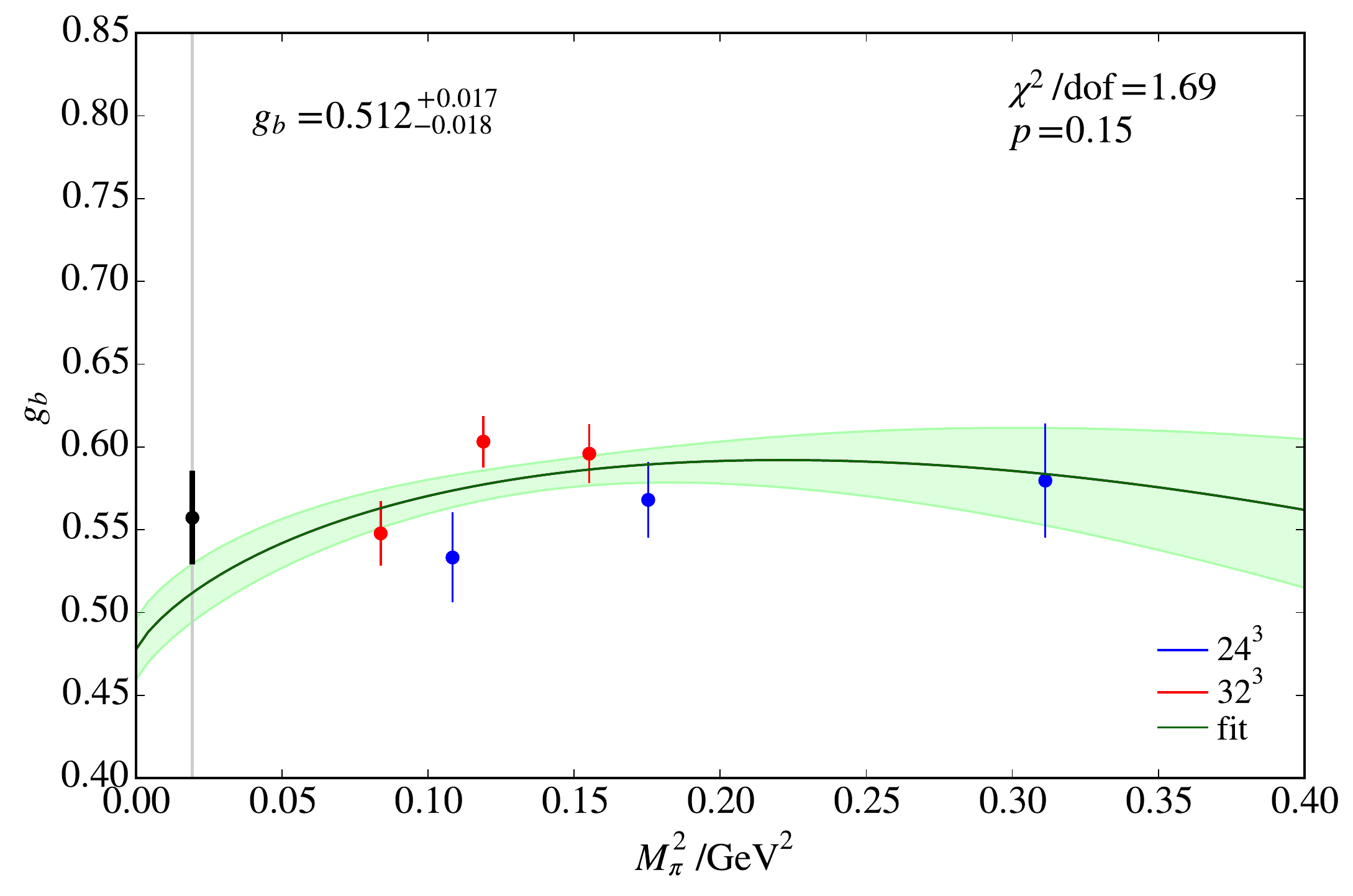}\\
% 2nd row
\includegraphics[width=0.496\textwidth ]{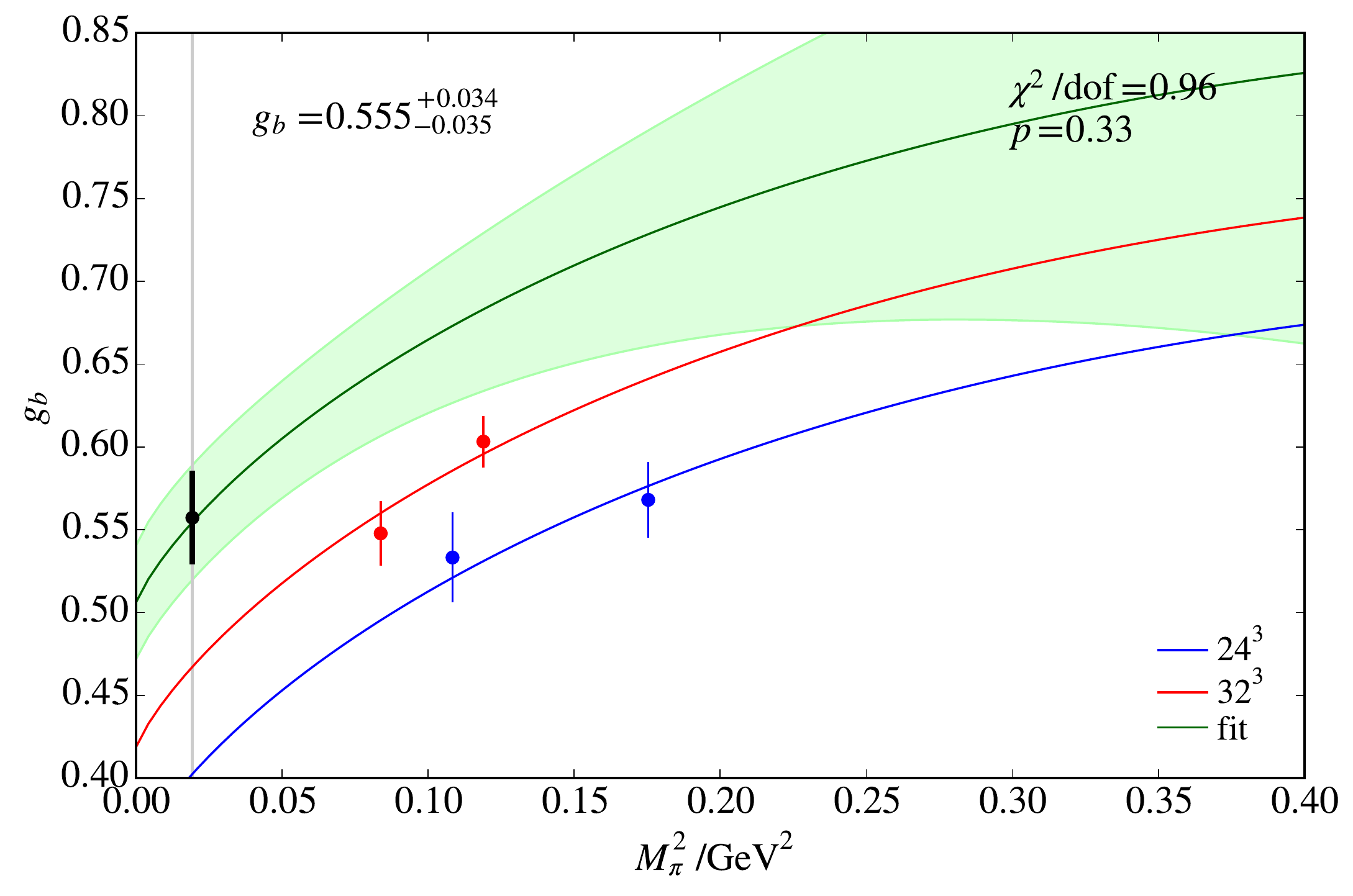}%
\hspace*{\fill}
\includegraphics[width=0.496\textwidth ]{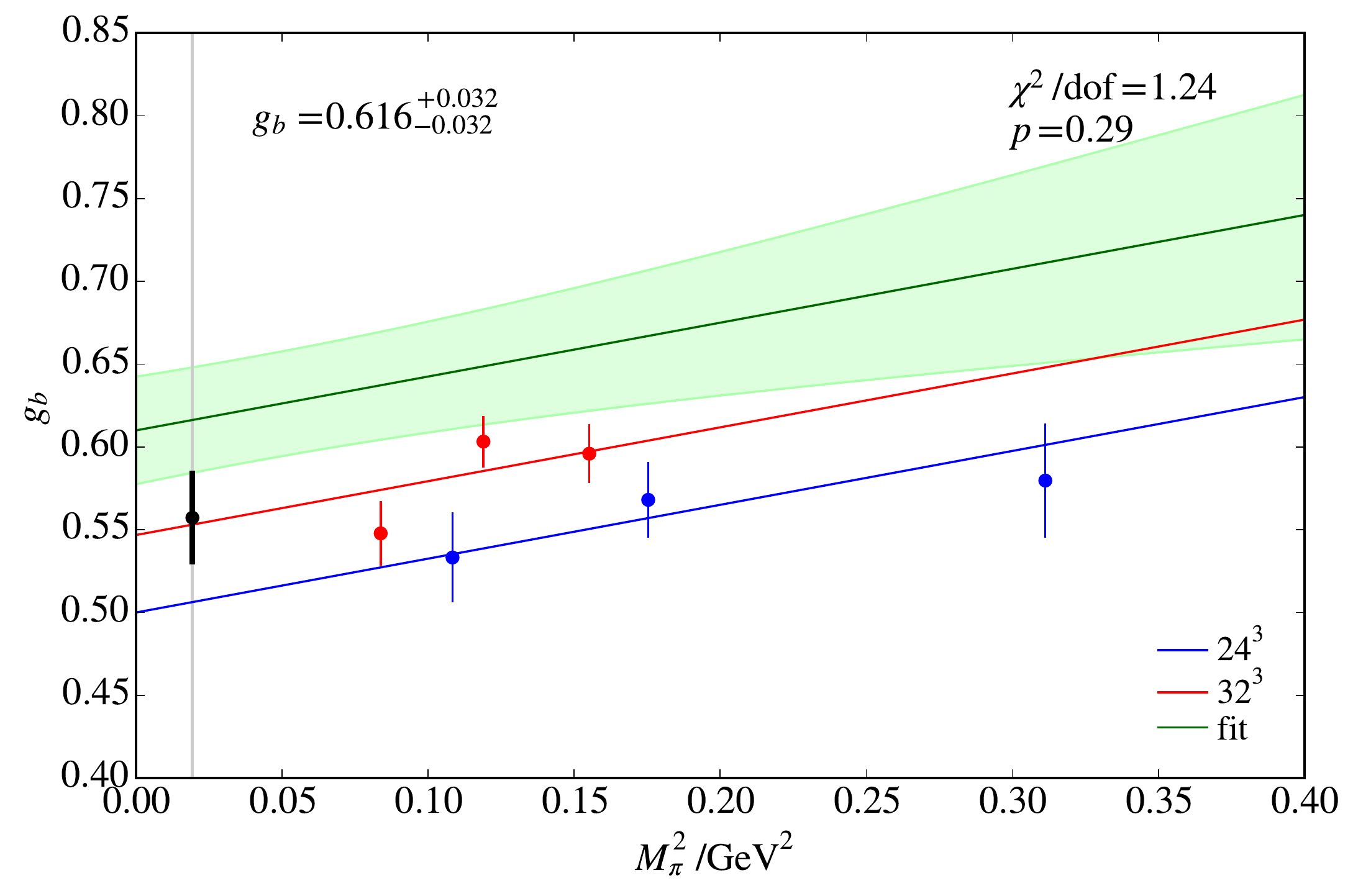}\\
% 3rd row
\includegraphics[width=0.496\textwidth ]{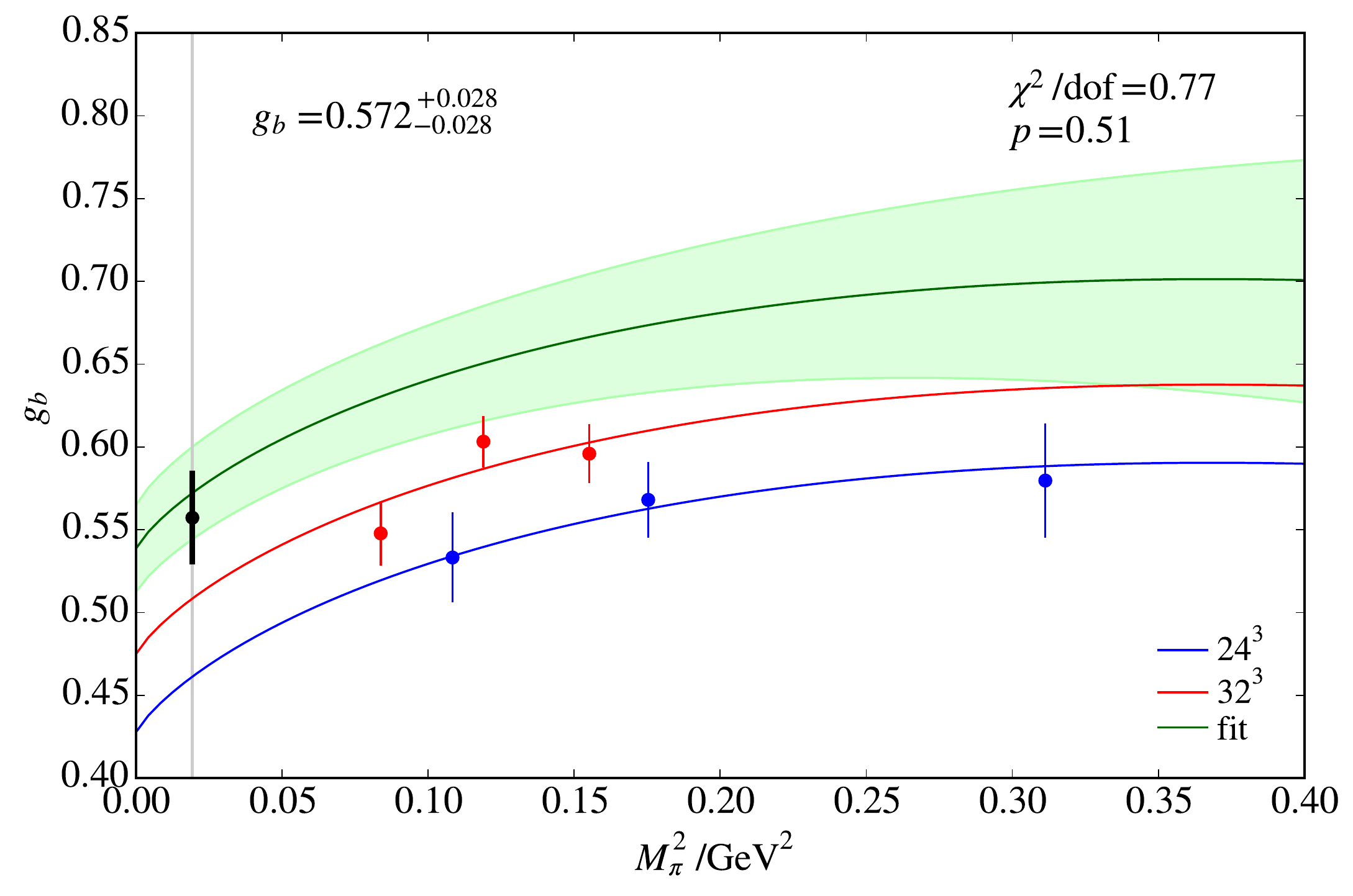}%
\hspace*{\fill}
\includegraphics[width=0.496\textwidth ]{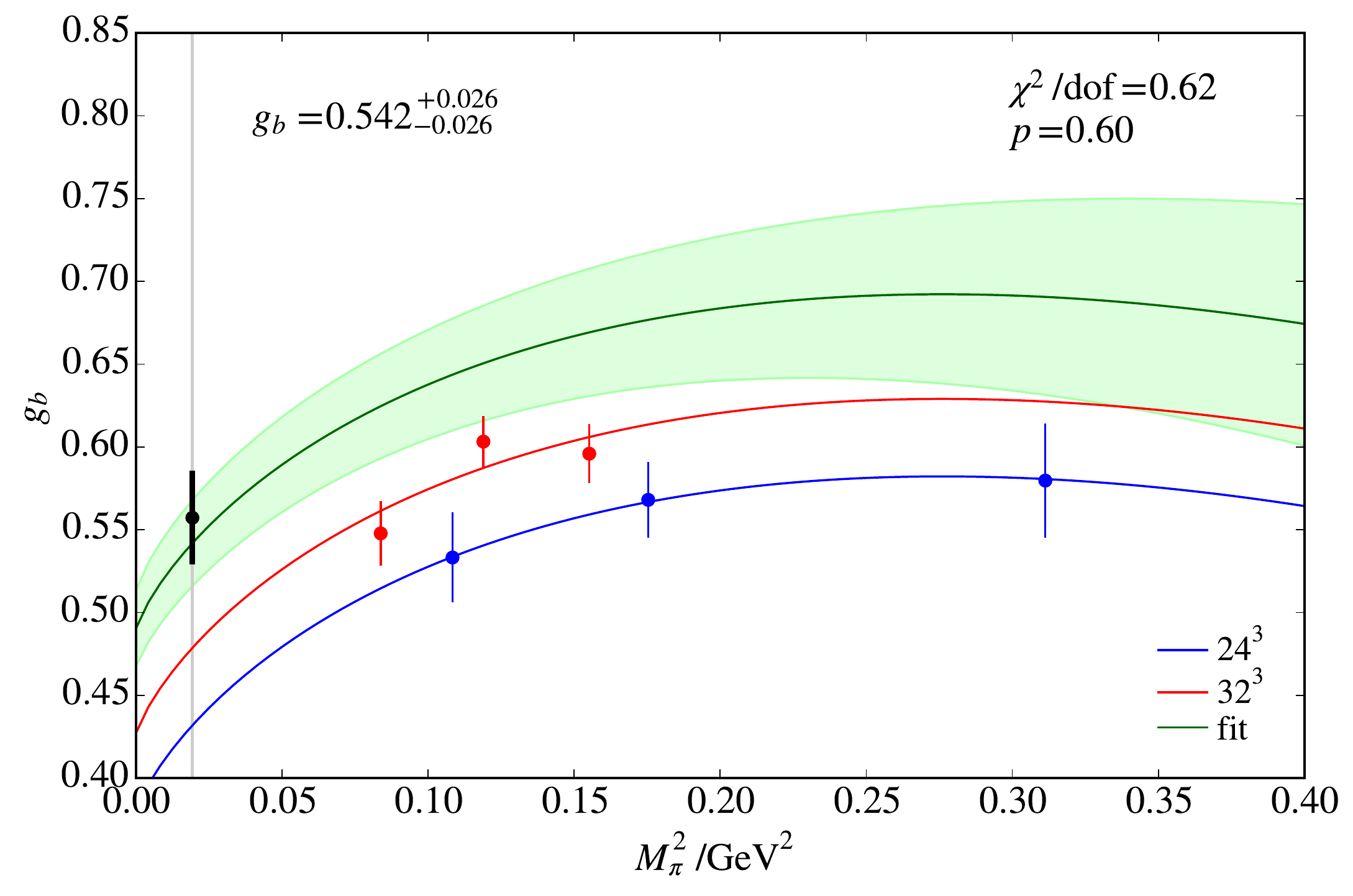}
\end{center}
\caption{Variations of chiral fits. Top left: $32^2$ data points only;
  top right: data points from both lattices with no $a^2$ term. Center
  left: heaviest masses dropped; center right: fit to a function
  linear in $M_\pi^2$. Bottom left: replacing $f_\pi$ with
  $f_K=156.1\mev$; bottom right: replacing $f_\pi$ with $f_0=112\mev$.
  In each plot, the result of the preferred fit from
  Fig.~\ref{chiral_extrap} is shown as the black point, with error, on
  the vertical line at the physical pion mass.}
\label{fig:fitvariations}
\end{figure*}

Our chiral extrapolation relies on NLO SU(2) \hmchipt{} with pion
masses above $400\mev$. Therefore we may be using the theory beyond
its range of applicability and we are certainly omitting higher order
terms in the chiral expansion. To estimate the uncertainty this
introduces we consider a range of possible fits. First, we consider
the effect of neglecting the heaviest mass from each ensemble (center
left plot in Fig.~\ref{fig:fitvariations}). This alters the form of
the fit dramatically but does not significantly change the final
result. In the bottom row of Fig.~\ref{fig:fitvariations} we replace
$f_\pi$ in the coefficient of the NLO chiral logarithms with
$f_K=156.1\mev$~\cite{Beringer2012} or with $f_0=112\mev$ in the SU(2)
chiral limit from the RBC/UKQCD light pseudoscalar meson
analysis~\cite{Aoki2010a}. This changes the relative size of NLO and
NNLO and higher-order terms in the chiral expansion. Finally, we note
that our data does not show any strong evidence of chiral log
curvature, presumably because our lightest data point corresponds to
$M_\pi \approx 289\mev$ and is still rather heavy. We therefore
consider an analytic fit, shown in the centre right plot of
Fig.~\ref{fig:fitvariations}, where we extrapolate linearly in
$M_\pi^2$. Of these variations, the largest difference from our
central value for $g_b$ is from the linear fit in $M_\pi^2$ and $a^2$.
This value is larger than our full chiral-continuum fit by $10.6\%$.
Because the chiral and continuum extrapolations are treated together
in our fitting procedure, however, discretization and chiral
extrapolation errors cannot be fully disentangled. In
section~\ref{sec:lqg-discretization-errors} below we consider
light-quark and gluon discretization errors, estimating a systematic
error of $11.5\%$. This is the largest deviation seen in the
chiral-continuum fits in Fig.~\ref{fig:fitvariations} and is therefore
the error we take for the combined chiral and continuum extrapolation.

\subsection{Lattice-scale uncertainty}

The coupling $g_b$ is a dimensionless number calculated from ratios of
correlators, so it should have only a mild dependence on the physical
value of the lattice spacing. However, variations in $a$ affect the
chiral and continuum extrapolations. We estimate the error in $g_b$
due to the lattice-spacing uncertainty by varying the $24^3$ and
$32^3$ lattice spacings by their quoted (statistical plus systematic)
uncertainties, $\sigma_{24}$ and $\sigma_{32}$~\cite{Aoki2010a}, one
at a time whilst keeping the other fixed. Shifting the lattice spacing
on the finest ensemble changes $g_b$ by $0.7\%$, and on the coarser
ensemble $g_b$ changes by $0.6\%$. Therefore ascribing an error of
$0.9\%$ (the sum in quadrature) to this source of uncertainty seems a
conservative estimate.

\subsection{Unphysical sea strange-quark mass}
\label{sec:strange}

Our simulation is performed with a sea strange-quark mass that differs
from the physical value by approximately $10\%$. To investigate the
effect of the sea strange-quark mass on $g_b$ we use results
from~\cite{Detmold2011} for the NLO axial current matrix element in
partially quenched \hmchipt. This allows us to evaluate the expression
with different valence and sea strange-quark masses. The NLO matrix
element has four different contributions, coming from so called sunset
diagrams, wave-function renormalization, tadpole diagrams and the NLO
analytic terms. We have calculated the effect of a $10\%$ change in
the sea strange-quark mass in the loop diagrams, assuming the values
of the low-energy constants obtained from our preferred chiral fit, on
the value of the coupling $g_b$. We find a change in $g_b$ of $1.5\%$.
This result is numerically consistent with the effect of the strange
sea-quark mass on the pion decay constant observed by the RBC/UKQCD
collaboration in~\cite{Aoki2010a}. Therefore we ascribe an error of
$1.5\%$ in $g_b$ due to the unphysical strange-quark mass.

\subsection{RHQ parameter uncertainties}
\label{subsec:RHQ-param-uncertainties}

\subsubsection{Statistical}

To test the dependence of $g_b$ on the uncertainties in the tuned RHQ
parameters we calculate the coupling on the $24^3$ $m_la=0.005$
ensemble using the full ``box'' of RHQ parameters used to interpolate
to the tuned values:
\begin{equation}
  \label{eq:parambox}
  \left[\begin{array}{c}m_0a\\c_p\\\zeta\end{array}\right],
  \left[\begin{array}{c}m_0a\pm\sigma_{m_0a}\\c_p\\\zeta\end{array}
  \right],
  \left[\begin{array}{c}m_0a\\c_p\pm\sigma_{c_p}\\\zeta\end{array}
  \right],
  \left[\begin{array}{c}m_0a\\c_p\\\zeta\pm\sigma_{\zeta}\end{array}
  \right].
\end{equation}
For our $24^3$ ensembles, the box parameters are given by
\begin{equation}
\begin{aligned}
(m_0a,c_p,\zeta) &= (8.40,5.80,3.20),\\
(\sigma_{m_0a},\sigma_{c_p},\sigma_\zeta) &= (0.15,0.45,0.30).
\end{aligned}
\end{equation}
We then linearly interpolate $g_b$ to the point of the tuned
parameters. By following this procedure underneath the jackknife we
can propagate the statistical errors from parameter tuning through to
$g_b$. Comparison of this determination to the result calculated
directly using the tuned values of the parameters gives a measure of
how sensitive $g_b$ is to the uncertainties arising from the tuning.
We find that the central values differ by $0.01\%$ and the errors
agree to two significant figures. In the context of the overall
uncertainty this can be considered negligible.

\begin{figure}
 \begin{center}
  \includegraphics[width=\linewidth ]{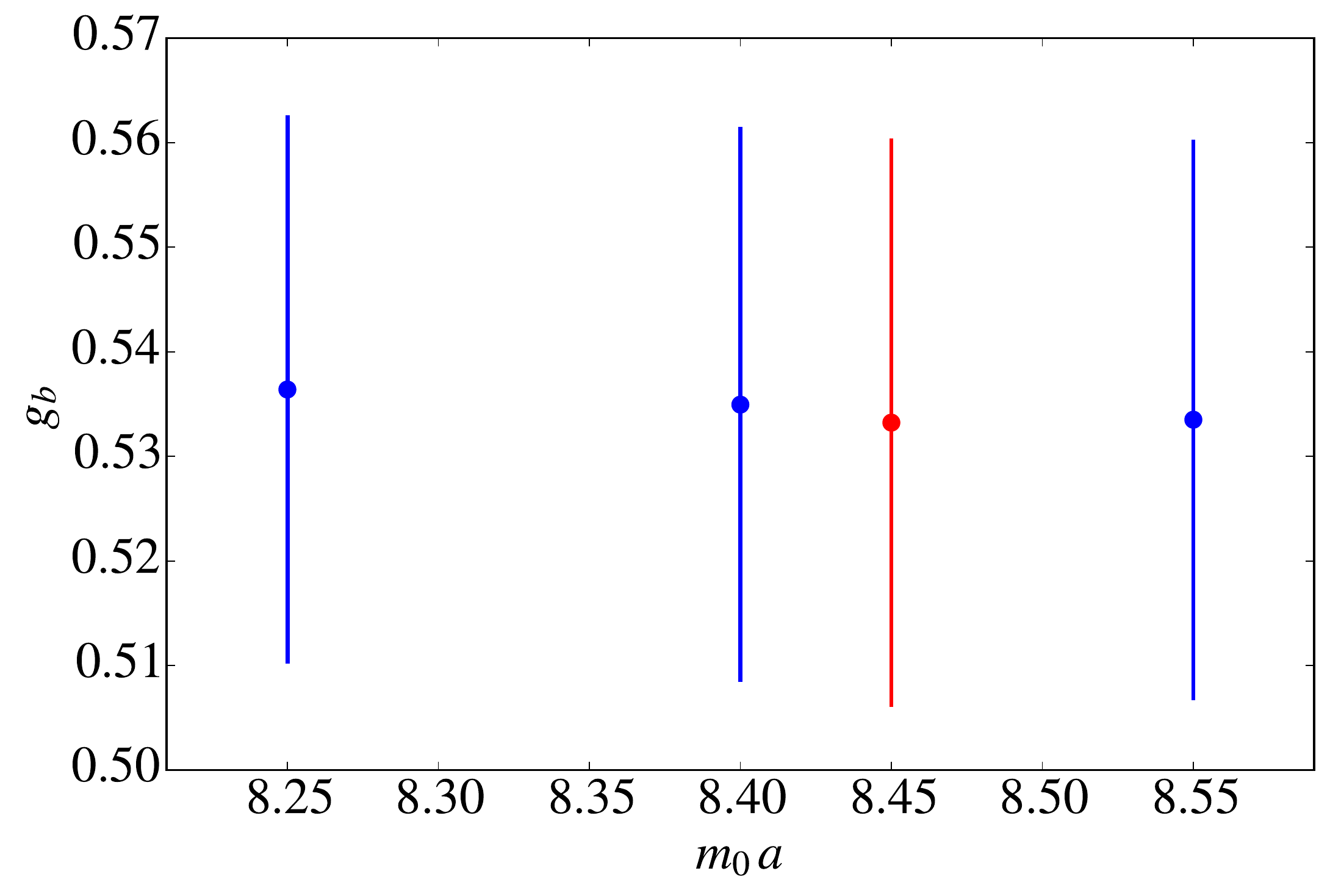}\\
  \includegraphics[width=\linewidth ]{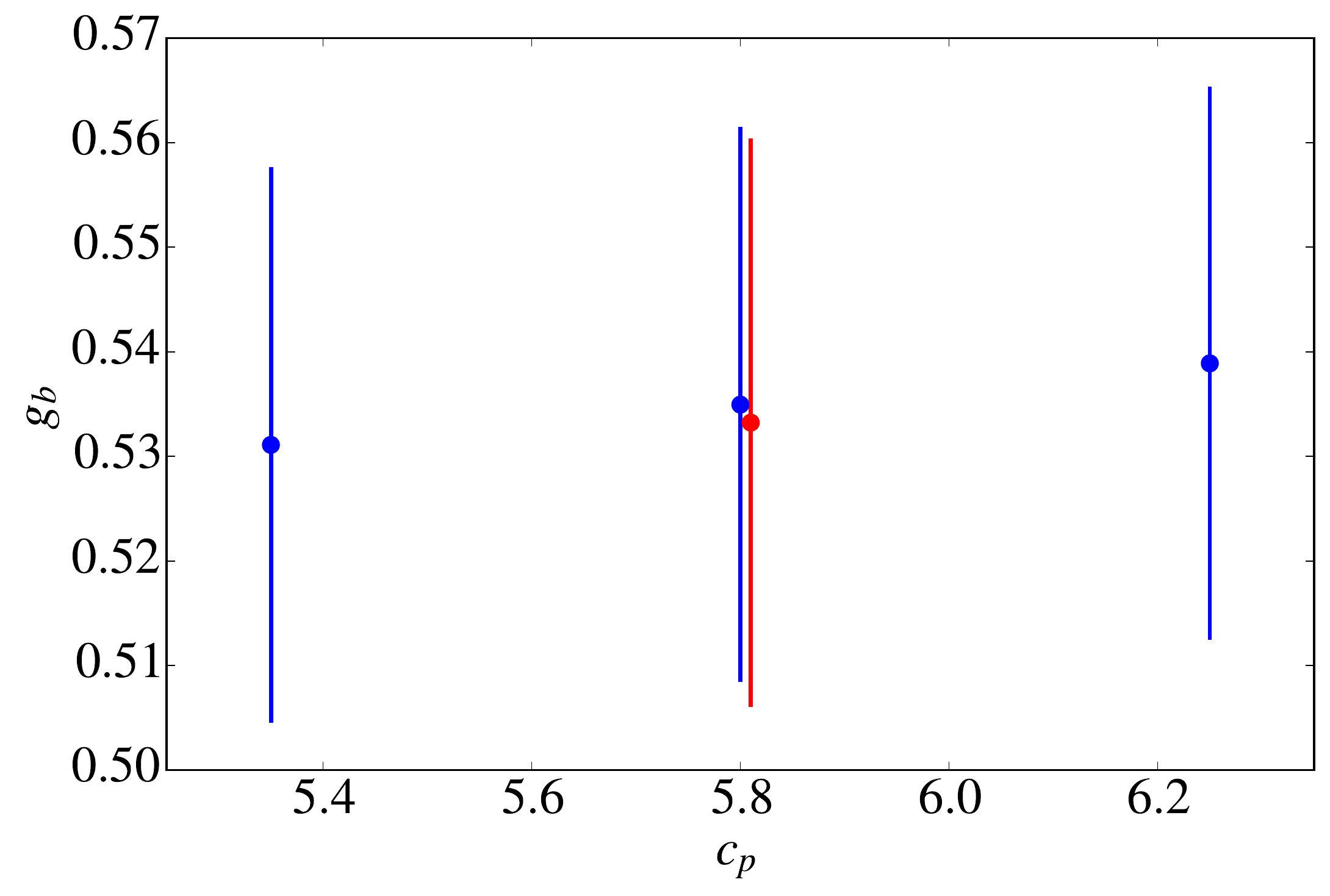}\\
  \includegraphics[width=\linewidth ]{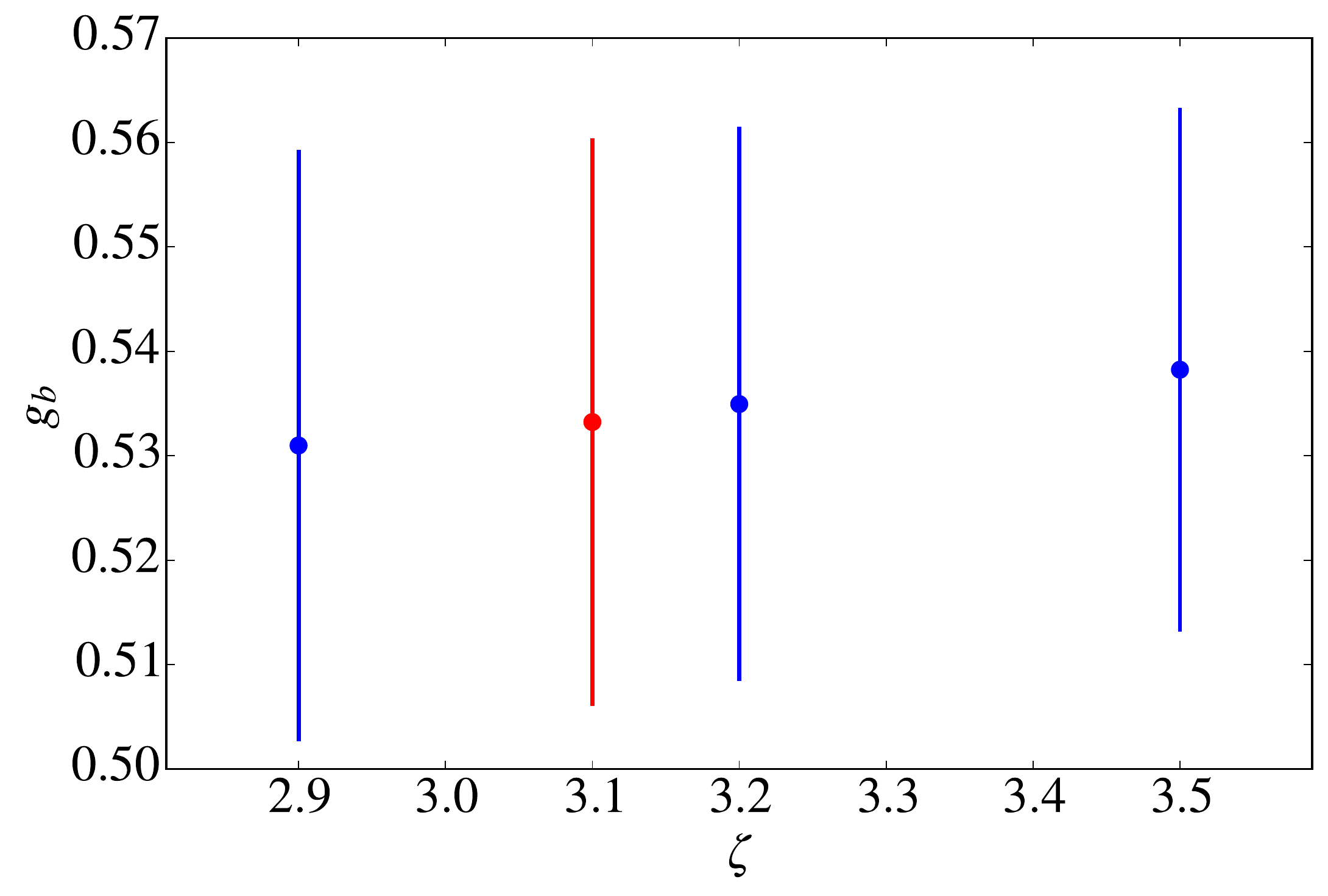}
 \end{center}
 \caption{$g_b$ calculated for the sets of RHQ parameters used to
   define the parameter ``box'' on the $24^3$ $m_la=0.005$ ensemble.
   The blue points are the results for the box parameter choices and
   the red point shows $g_b$ calculated directly at the tuned
   parameter values.}
 \label{fig:RHQ_err_stat}
\end{figure}

Figure~\ref{fig:RHQ_err_stat} shows $g_b$ calculated on the seven sets
of parameters indicated in Eq.~\eqref{eq:parambox} for the $24^3$
$m_la=0.005$ ensemble.

\subsubsection{Systematic}
We also consider the effect on $g_b$ of systematic uncertainties in
the RHQ parameters. These are estimated in Ref.~\cite{Aoki:2012xaa}
and given in Table~\ref{tab:RHQparams}. The three significant
contributors are heavy-quark discretization effects, uncertainty in
the lattice spacing, and uncertainty from the experimental inputs. To
determine the sensitivity of $g_b$ to these uncertainties we use the
calculation on the box of parameters, Eq.~\eqref{eq:parambox},
described in the previous subsection. We assume a linear dependence of
$g_b$ on the RHQ parameters for small shifts, then shift one parameter
at a time by each systematic uncertainty and take the overall error as
the effect of each of these shifts added in quadrature. The combined
effect, shown in Table~\ref{tab:rhq_sys_err}, is an error of $1.5\%$
in $g_b$.

\begin{table}
\caption{The effect of systematic uncertainties in the RHQ parameters
  on $g_b$. Each parameter was shifted by the uncertainty from each
  source and the effect on $g_b$ calculated by assuming $g_b$ depends
  linearly on the parameters.}
\begin{ruledtabular}
\begin{tabular}{lcccc}
                    & \multicolumn{4}{c}{\% error from parameter}\\
\cline{2-5}
Source              & $m_0a$  & $c_p$  & $\zeta$ & Total\\\hline
HQ discretization   &  0.25   & 0.65   & 0.30    & 0.76\\
Lattice scale       &  0.97   & 0.65   & 0.24    & 1.19\\
Experimental inputs &  0.14   & 0.33   & 0       & 0.35\\\hline
Total               &  1.01   & 0.98   & 0.38    & 1.46
\end{tabular}
\end{ruledtabular}
\label{tab:rhq_sys_err}
\end{table}

\subsection{Discretization errors}

\subsubsection{Heavy-quark discretization errors} 

We estimate heavy-quark discretization errors using an effective field
theory approach~\cite{symanzik1983continuum,El-Khadra1996a,Oktay2008}
in which both our lattice theory and QCD are described by effective
continuum Lagrangians built from the same operators and errors stem
from mismatches between the short-distance coefficients of
higher-dimension operators in the two effective theories. Oktay and
Kronfeld~\cite{Oktay2008} have catalogued the relevant operators and
calculated the mismatch coefficients at tree level.

Because we are evaluating a matrix element of the light-quark
axial-vector current, heavy-quark discretization errors stem from
mismatches in higher-dimension operators in the heavy-quark action
which correct the $B$ and $B^*$ meson masses. We expect these effects
to be negligible. From our tuning procedure~\cite{Aoki:2012xaa} we can
relate changes in the meson masses to changes in the RHQ parameters
$m_0a$, $c_p$ and $\zeta$, while in
section~\ref{subsec:RHQ-param-uncertainties} below, we relate changes
in the RHQ parameters to changes in $g_b$. Hence we can estimate the
effect of errors in the meson masses on $g_b$.

In Appendix~C of~\cite{Aoki:2012xaa}, we estimated the heavy-quark
discretization error on the spin-averaged $B_s$ meson mass as
$0.05\%$. Also in~\cite{Aoki:2012xaa}, that spin-averaged mass was
most sensitive to variations in $m_0a$, with a $0.05\%$ shift
corresponding to a change of around $0.02$ in $m_0a$ on the $24^3$
$m_la=0.005$ ensemble. From
section~\ref{subsec:RHQ-param-uncertainties}, shifting $m_0a$ by the
halfwidth of our tuning ``box'' changes $g_b$ by no more than $1.5\%$.
For the $24^3$ $m_la=0.005$ ensemble, this shift in $m_0a$ is $0.15$
and hence we expect a heavy-quark discretization error on $g_b$ of no
more than $(0.02/0.15)\times 1.5\% = 0.2\%$, which is negligible
compared to our overall uncertainty.

\subsubsection{Light-quark and gluon discretization errors}
\label{sec:lqg-discretization-errors}

Leading discretization errors from the domain-wall light-quark action
and the Iwasaki gauge action are both $O(a^2)$ and are included as an
$a^2$ term in the combined chiral-continuum extrapolation. However the
data is also compatible within errors with assuming no lattice-spacing
dependence; a fit with no $a^2$ term also yields an acceptable, albeit
larger, $\chi^2$/dof. The top row of Fig.~\ref{fig:fitvariations}
shows chiral fits to the data without an $a^2$ term. To estimate the
systematic errors coming from the continuum extrapolation we use the
difference of $11.5\%$ in $g_b$ between a fit to our finest data set
($a\approx0.086$fm) and the $a^2$ extrapolation using both lattice
spacings. This is the largest effect in all variations of our chiral
and continuum extrapolations and is therefore the value appearing in
Table~\ref{tab:errors} for the combined chiral and continuum
extrapolation uncertainty.

\subsection{Finite-volume effects}

We expect that finite-volume effects are small since there are no
propagating light particles in the simulated system. To estimate their
size we compare the value of $g_b$ obtained from an NLO heavy-meson
$\chi$PT fit to our data, with and without finite volume effects
included. We compare the finite and infinite-volume fit result at all
of our simulated pion-mass values. The largest finite-volume
correction, which occurs for our lightest pion mass, is $\lesssim
1\%$, so we take $1\%$ as the finite-volume error in our calculation of
$g_b$.

\section{Conclusions}\label{sec:conclusions}

The sum in quadrature of all the systematic errors described in
section~\ref{sec:sys} gives a total systematic uncertainty of 12\%.
Our final error budget is given in Table~\ref{tab:errors} and our
final value of the coupling $g_b$ including statistical and systematic
errors is
\begin{equation}
  g_b = 0.56(3)_\mathrm{stat}(7)_\mathrm{sys}
\end{equation}
 
\begin{table}
  \caption{Error budget for systematic and statistical errors.}
     \begin{ruledtabular}
       \begin{tabular}[c]{lr}
         Statistical errors & 4.8\% \\
         \hline
         Chiral and continuum extrapolation & 11.5\%\\
         Lattice scale uncertainty & 0.9\% \\
         Finite volume effects & 1.0\% \\
         RHQ parameter uncertainties & 1.5\%\\
         Unphysical sea strange-quark mass & 1.5\%\\
         \hline
         Systematic errors total & 11.8\% \\
       \end{tabular}
    \end{ruledtabular}
  \label{tab:errors}
\end{table}

\begin{figure}
  \begin{center}
  \includegraphics[width=\linewidth ]{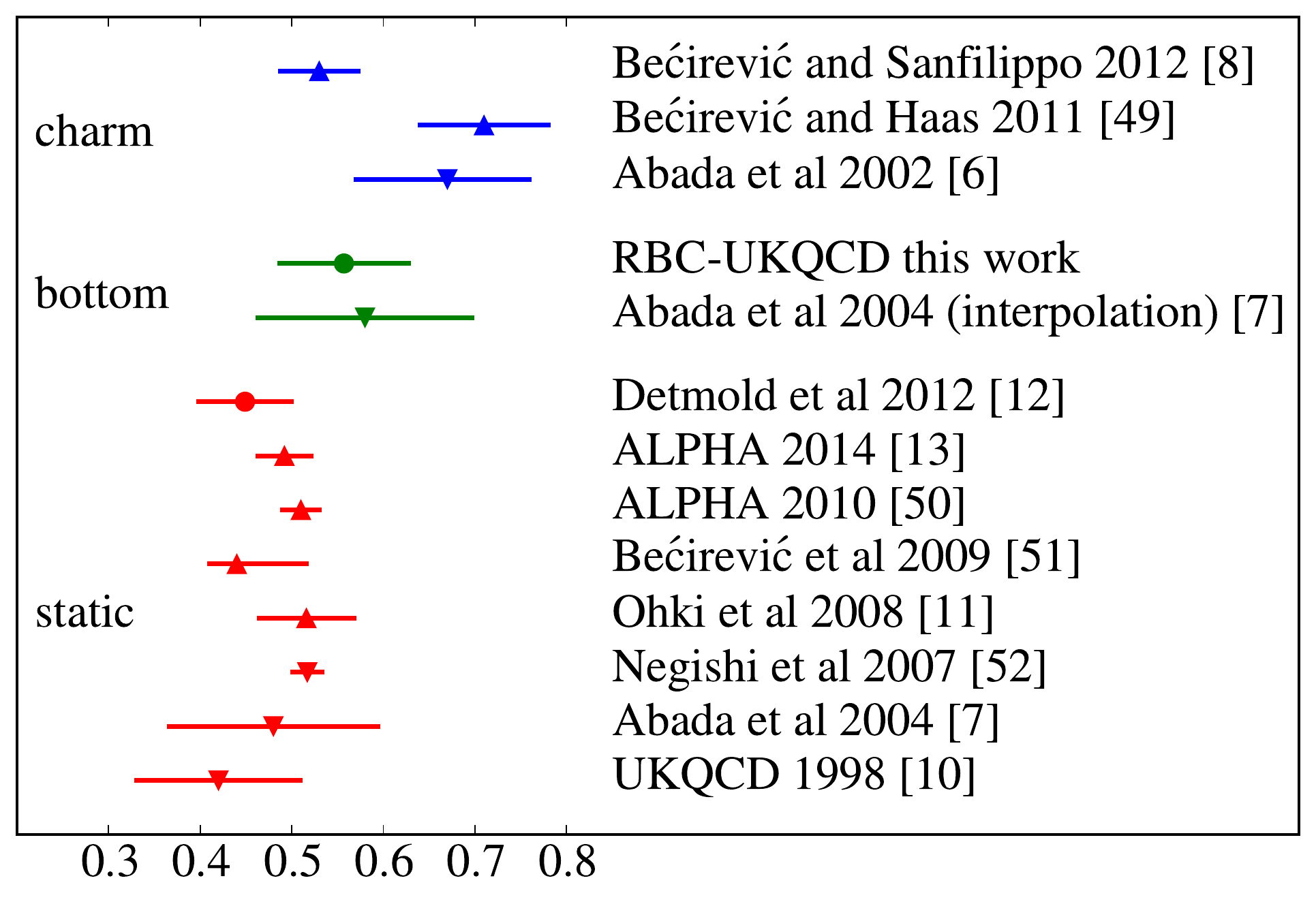}
  \end{center}
  \caption{Comparison of quenched (downward triangles), $N_f = 2$
    (upward triangles) and $N_f = 2+1$ (circles) calculations of the
    couplings $g_c$, $g_b$ and $\hat
    g$~\cite{Becirevic2012,Becirevic2009b,Abada2002,Becirevic2004,Detmold2012,Bernardoni:2014kla,Bulava2010,Becirevic2009,Ohki2008,Negishi:2006sc,DeDivitiis1998}.
    Error bars represent the sum in quadrature of all quoted errors
    (statistical and systematic).}
  \label{fig:compare}
\end{figure}

Our calculation is the first directly at the physical $b$-quark mass,
and has a complete systematic error budget. Fig.~\ref{fig:compare}
compares our result to earlier dynamical calculations at the
charm-quark mass and in the static limit. The dependence of $g$ on the
value of the heavy-quark mass is mild, and our result lies in the
region that would be expected from interpolating between the charm-
and infinite-mass determinations. Our result is compatible with the
experimental value $g_c^\mathrm{exp}=0.570\pm0.004\pm0.005$ extracted
from the natural linewidth of the transition $D^*(2010)^+ \to
D^0\pi^+$ by the BaBar Collaboration in~\cite{Lees:2013uxa}. This
further suggests that $1/m_Q^n$ corrections to the coupling $g$ are
small. Our result has been used by the RBC/UKQCD collaboration in the
chiral extrapolations of numerical lattice data for the $B$-meson
leptonic decay constants~\cite{Witzel:2012pr,Christ:2014uea} and
$B\to\pi\ell\nu$ and $B_s\to K\ell\nu$ semileptonic form
factors~\cite{Kawanai:2012id,Flynn:2015mha}.

\begin{acknowledgments}
BS was supported by EPSRC Doctoral Training Centre Grant EP/G03690X/1.
JMF and CTS were supported by STFC Grants ST/J000396/1 and
ST/L000296/1. We acknowledge the use of the Iridis High Performance
Computing Facility and associated support services at the University
of Southampton, USQCD resources at Fermilab, in part funded by the
Office of Science of the U.S.\ Department of Energy, as well as
computers at Brookhaven National Laboratory and Columbia University.
Fermilab is operated by Fermi Research Alliance, LLC, under Contract
No. DE-AC02-07CH11359 with the U.S.\ Department of Energy.
\end{acknowledgments}

\bibliography{../bbpi2}
\end{document}